\documentclass[journal]{IEEEtran}
\usepackage{amsmath,amsfonts,amssymb}
\usepackage[ruled,vlined,linesnumbered]{algorithm2e}
\usepackage[table,xcdraw]{xcolor}
\usepackage[colorlinks,linkcolor=black,citecolor=blue]{hyperref}
\SetKwComment{Comment}{\color{blue}\hfill// }{}
\usepackage{array}
\usepackage{float}
\usepackage{graphicx}
\usepackage{booktabs}
\usepackage{multirow}
\usepackage{makecell}
\usepackage{subfig}
\usepackage{orcidlink}
\usepackage{marvosym}
\usepackage{textcomp}
\usepackage{stfloats}
\usepackage{url}
\usepackage{verbatim}
\usepackage{graphicx}
\usepackage{tikz}
\usepackage{threeparttable}

\newcommand{\orcid}[1]{\href{https://orcid.org/#1}{\textcolor[HTML]{A6CE39}{\aiOrcid}}}

\def\BibTeX{{\rm B\kern-.05em{\sc i\kern-.025em b}\kern-.08em
    T\kern-.1667em\lower.7ex\hbox{E}\kern-.125emX}}
\usepackage{balance}

\begin{document}
\title{A Scenario-Oriented Survey of Federated Recommender Systems: Techniques, Challenges, and Future Directions}

\author{Yunqi Mi\orcidlink{0009-0009-8624-4818}\textsuperscript{\dag},
Jiakui Shen\orcidlink{0009-0009-4926-3809}\textsuperscript{\dag},
Guoshuai Zhao\orcidlink{0000-0003-4392-8450}\textsuperscript{\Letter}, ~\IEEEmembership{Member,~IEEE,}
Jialie Shen\orcidlink{0000-0002-4560-8509}, ~\IEEEmembership{Senior Member,~IEEE,} and Xueming Qian\orcidlink{0000-0002-3173-6307}

\thanks{\textsuperscript{\dag} These authors contributed equally to this research.}
\thanks{\textsuperscript{\Letter} Corresponding author.}

\thanks{Yunqi Mi, Jiakui Shen, and Guoshuai Zhao are with the School of Software Engineering, Xi’an Jiaotong University, Xi’an
710049, China (e-mail: miyunqi@stu.xjtu.edu.cn; shenjiakui@stu.xjtu.edu.cn;
guoshuai.zhao@xjtu.edu.cn}
\thanks{Jialie Shen is with City, University of London, U.K.(e-mail: jerry.shen@city.ac.uk).}
\thanks{Xueming Qian is with the Ministry of Education Key Laboratory for Intelligent Networks and Network Security, the School of Information and Communication Engineering, and SMILES LAB, Xi’an Jiaotong University, Xi’an
710049, China (e-mail: qianxm@mail.xjtu.edu.cn).}}

\maketitle

\begin{abstract}

Extending recommender systems to federated learning (FL) frameworks to protect the privacy of users or platforms while making recommendations has recently gained widespread attention of academia.
This is due to the natural coupling of recommender systems and federated learning architectures: the data originates from distributed clients (mostly mobile devices held by users), which are highly related to privacy.
In a centralized recommender system (CenRec), the central server collects clients' data, trains the model and provides the service. Whereas in federated recommender systems (FedRec), the step of data collecting is omitted, and the step of model training is offloaded to each client. The server only aggregates the model and other knowledge, thus avoiding client privacy leakage.
Some surveys of federated recommender systems discuss and analyze related work from the perspective of designing FL systems.
However, their utility drops by ignoring specific recommendation scenarios' unique characteristics and practical challenges.
For example, the statistical heterogeneity issue in cross-domain FedRec originates from the label drift of the data held by different platforms, which is mainly caused by the recommender itself but not the federated architecture.
Therefore, it should focus more on solving specific problems in real-world recommendation scenarios to encourage deployment FedRec.
To this end, this review comprehensively analyzes the coupling of recommender systems and federated learning from the perspective of recommendation researchers and practitioners. 
We establish a clear link between recommendation scenarios and FL frameworks, systematically analyzing scenario-specific approaches, practical challenges, and potential opportunities.
We aim to develop guidance for the real-world deployment of FedRec, bridging the gap between existing research and applications.
To facilitate subsequent exploration, we provide a continuously updated repository as:~\url{https://github.com/SmilesLab-XJTU/Survey-FedRec}.

\end{abstract}


\section{Introduction}

\begin{figure*}
	\centering
	\includegraphics[width=1\linewidth]{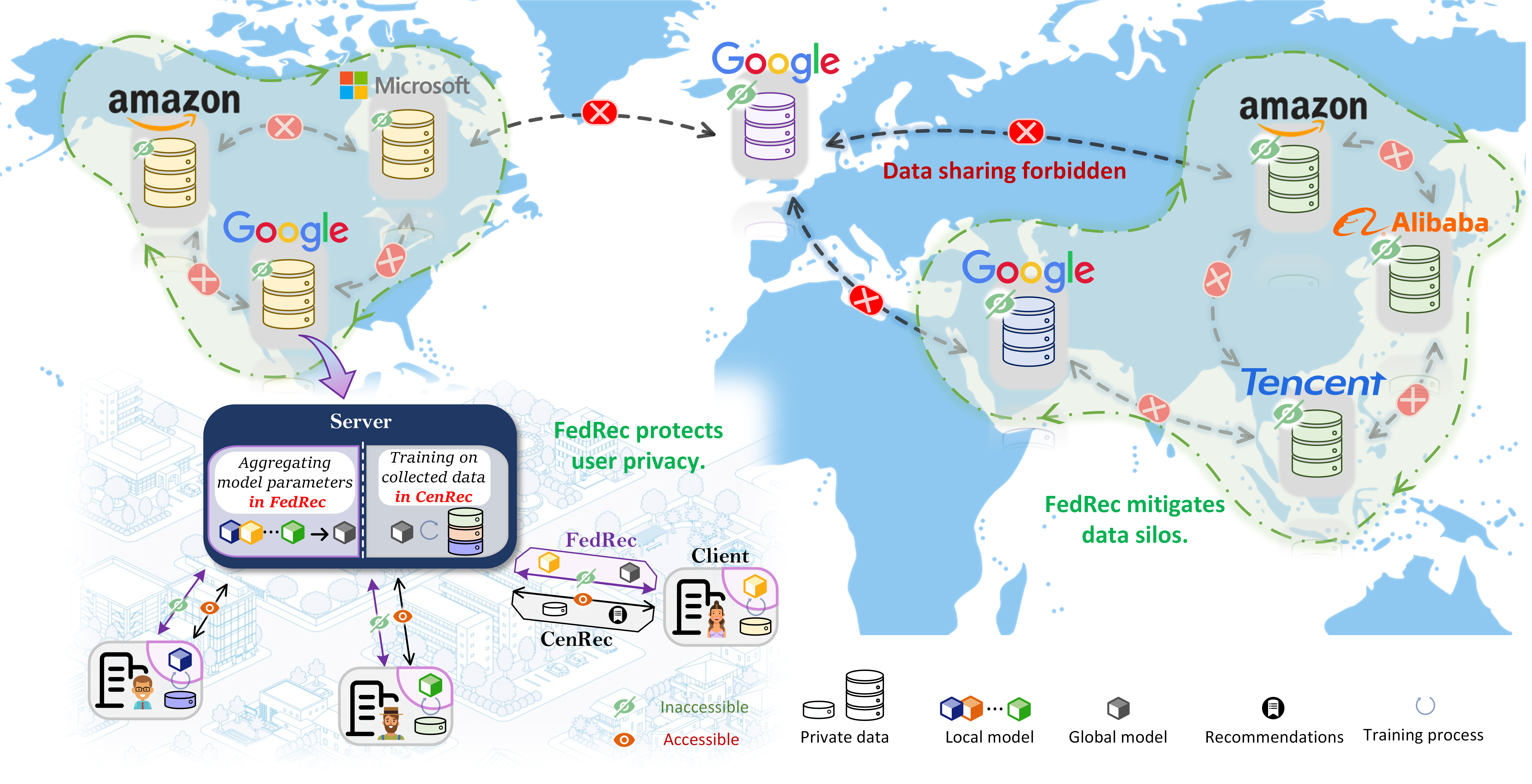}
	\caption{FedRec can be considered as a CenRec that offloads model training to the client and omits the data collection process. Therefore, deploying FedRec offers intuitive benefits for both individuals and enterprises. It not only efficiently protects user privacy but also alleviates data silos in a simple and feasible manner.}
	\label{fig:introduction}

\end{figure*}

Recommender systems (RS) rely heavily on user data, including behavioral records (\textit{e.g.}, ratings, clicks) and attributes (\textit{e.g.}, social connections, gender) \cite{intro1, intro2, intro3, jialie_survey}.
Recently, the growing demand for performance has led to the need to collect richer user information, raising widespread privacy concerns \cite{sun_survey}.
Besides, data across organizations and platforms is considered unauthorized assets which cannot share, leading to the phenomenon of data silos.
In addition, a series of data privacy regulations, such as the General Data Protection Regulation (GDPR\footnote{\url{https://gdpr-info.eu/}}), have been proposed to restrict direct access to data severely.
Therefore, since the simple and feasible approach adopted by federated learning (FL), which protects private data without reducing accuracy, combining RS with FL has become a promising technology, as shown in Fig.~\ref{fig:introduction}.

There is an obvious coupling between RS and FL: the recommendation data is held by distributed clients and is highly related to privacy.
In a centralized recommender system (CenRec), the central server collects these distributed data, centrally trains the model and makes suggestions for each client.
Whereas in a federated recommender system (FedRec), the process of collecting private data is omitted, and the central server only collects privacy-free parameters for aggregation. The training process and private data are always kept in clients, thus protecting the privacy.
In this way, FedRec can be regarded as the CenRec that offloads the model training process to the clients and then uses the server to aggregate the model parameters uploaded by the clients.

As a FL technology, FedRec typically adopts a client-server or peer-to-peer architecture.
The client-server architecture follows the above process, with data and interactions always kept within each client, while only models and other parameters are transmitted between the server and clients.
This architecture is relatively simple, the training and inference processes are more orderly and controllable, making it easy to deploy and maintain.
However, as the key to the entire system, this architecture will fail once the server fails or is hijacked.
Moreover, all clients need to communicate with the server, which causes the communication bottleneck on server when the number of clients is huge, affecting the training efficiency.
The peer-to-peer architecture removes the central server, each client directly exchanges and aggregates parameters with its neighbours.
Compared to the client-server architecture, it reduces the risk of a single point of failure and improves Fedrec's communication efficiency.
However, malicious clients can more easily propagate malicious model updates and contaminate the global model.

\begin{figure}
	\centering
\subfloat[Number of related publications per year.]{\includegraphics[width=3.53cm,height = 3.33cm]{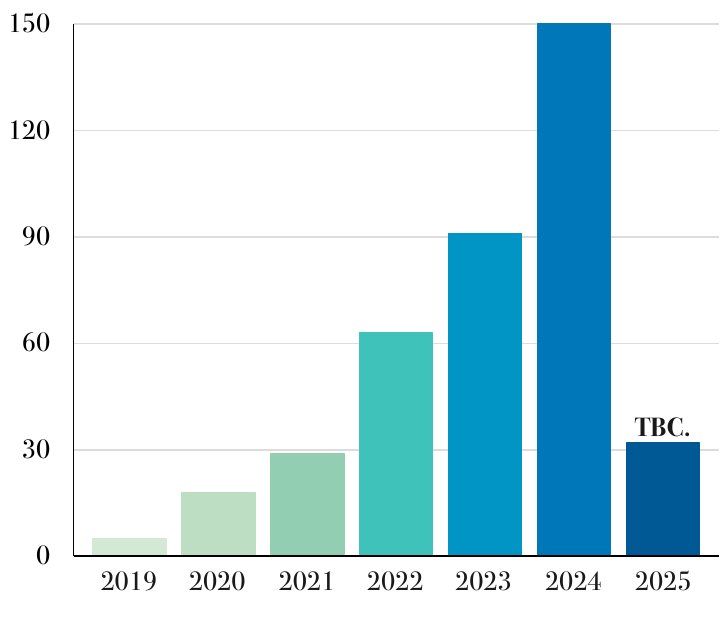}}
 \quad
  \subfloat[Comparison of publication numbers for fine-grained scenarios.]{\includegraphics[width=4.53cm,height = 3.33cm]{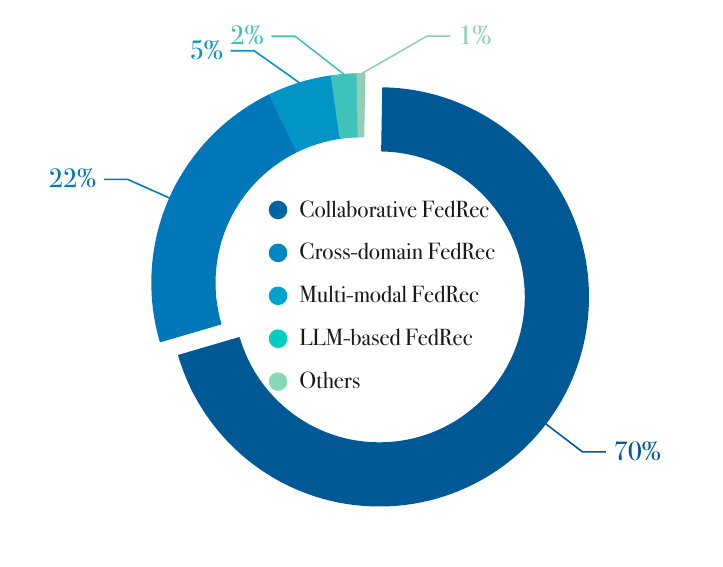}}
	\caption{The historical trend of FedRec publications from 2019 to Early-2025. It is evident that an increasing number of studies are focusing on FedRec, but they are limited to basic tasks, especially collaborative filtering.}
	\label{fig:statistic}
\end{figure}

Compared to other FL tasks, FedRec has the following three unique properties due to more complex scenarios and personalized sparser data on the client side:
1) \textbf{Large-scale restricted clients}. 
Although in some cases (\textit{e.g.}, cross-domain FedRec), the client usually denotes the platform and organization with large computing equipment.
In most scenarios, the client refers to light devices (\textit{e.g.}, smartphones, laptops) that contain only a single user.
These clients are significantly more limited in computing power and bandwidth than those business-type clients.
As a result, FedRec usually contains a huge number of clients \cite{HeteFedRec-yuan2024hetefedrec, F2PGNN-agrawal2024no, ReFRS-imran2023refrs, MetaMF-lin2020meta, IFedRec-zhang2024federated}, even up to the million level. 
2) \textbf{Sparse and unique interactions}. 
Due to the diversity of individuals, overall user interactions and profiles are unique, despite some similar behaviors or attributes among them \cite{FedRAP-lifederated, PFedRec-zhang2023dual, GPFedRec-zhang2024gpfedrec, FCF-ammad2019federated, FedMF-chai2020secure}. 
Besides, clients can only interact with a small portion of the ocean of items, leading to sparse interactions.
3) \textbf{Introduction of pseudo-interactions}.
Client interaction data is an important basis for user profiling and preference analysis.
Therefore, in order to protect real interactions, a widely adopted approach in FedRec is to sample a portion of uninteracted items to upload to the server as pseudo-interactions \cite{Semi-qu2023semi, FedPerGNN-wu2022federated}.
However, excessive pseudo-interactions decrease the model's recommendation performance and cause serious communication bottlenecks, while too few pseudo-interactions cause the exposion of client's preference.
Therefore, it is necessary to design lighter methods for FedRec to deal with extreme statistical heterogeneity among a large number of clients, and achieve a trade-off in recommendation performance, communication cost, and privacy protection.

\begin{table*}[t]
\caption{The review of FedRec research based on the proposed taxonomy.}
\centering
\renewcommand\arraystretch{1.5}
 \resizebox{\textwidth}{!}{
\begin{tabular}{|>{\centering\arraybackslash}m{2cm}|m{7cm}|m{6cm}|>{\raggedright\arraybackslash}m{5cm}|}
\hline
\rowcolor[HTML]{EFEFEF} 
{\color[HTML]{000000} \textbf{Category}} & \multicolumn{1}{c|}{\cellcolor[HTML]{EFEFEF}{\color[HTML]{000000} \textbf{Mechanism}}} & \multicolumn{1}{c|}{\cellcolor[HTML]{EFEFEF}{\color[HTML]{000000} \textbf{Core Idea}}} & \multicolumn{1}{c|}{\cellcolor[HTML]{EFEFEF}{\color[HTML]{000000} \textbf{Practical Use Case}}}\\ \hline
                        
\hline

Collaborative FedRec  &
FCF \cite{FCF-ammad2019federated}, FedMF \cite{FedMF-chai2020secure}, FedRec \cite{FedRec-lin2020fedrec}, FedRec++ \cite{FedRec++-liang2021fedrec++}, SP-FCF \cite{SP-FCF-StrongPrivacyCF-minto2021stronger}, FR-FMSS \cite{FR-FMSS-lin2021fr}, FedNCF \cite{FedNCF-perifanis2022federated}, HFSA \cite{HFSA}, LIBERATE \cite{LIBERATE}, HN3S \cite{HN3S}, FRU \cite{FRU-yuan2023federated}, FedIS \cite{FedIS}, DP-PrivRec \cite{DP-PrivFec-wang2022fast}, CLOUD \cite{CLOUD}, CFRU\cite{CFRU}, DGFedRS \cite{DGFedRS}, OVFR \cite{OVFR}, FedFast \cite{FedFast-muhammadFedFastGoingAverage2020a}, FedNewsRec \cite{FedNewsRec-qi2020privacy}, RF$^2$ \cite{RF2-maeng2022towards}, PFedRec \cite{PFedRec-zhang2023dual}, FedRAP \cite{FedRAP-lifederated}, GPFedRec \cite{GPFedRec-zhang2024gpfedrec}, PerFedRec \cite{PerFedRec-luo2022personalized}, PerFedRec++ \cite{PerFedRec++-luo2024perfedrec++}, PrefFedPOI \cite{PrefFedRec-zhang2023fine}, CPF-GCN \cite{CPF-GCN}, PO-FCF \cite{PO-FCF-khan2021payload}, CoLR \cite{CoLR-nguyen2024towards}, F$^2$PGNN \cite{F2PGNN-agrawal2024no}, HeteFedRec \cite{HeteFedRec-yuan2024hetefedrec}, P4GCN \cite{P4GCN}, LightFR \cite{LightFR-zhang2023lightfr}, DMFR \cite{DMFR}, ReFRS \cite{ReFRS-imran2023refrs}, MetaMF \cite{MetaMF-lin2020meta}, IFedRec \cite{IFedRec-zhang2024federated}, Efficient-FedRec \cite{Efficient-FedRe-yi2021efficient}, PTF-FedRec \cite{PTF-FedRec-yuan2024hide}, FR-SCU \cite{FR-SCU}, FedPerGNN \cite{FedPerGNN-wu2022federated}, P-GCN \cite{P-GCN-hu2023privacy}, VerFedGNN \cite{VerFedGNN-mai2023vertical}, SemiDFEGL \cite{Semi-qu2023semi}, CDCGNNFed \cite{CDCGNNFed-qu2024towards}, FedHGNN \cite{FedHGNN-yan2024federated}, DeFedGCN \cite{DeFedGCN}, A-ra \cite{A-ra-rong2022poisoning}, FedRecAttack \cite{FedRecAttack-rong2022fedrecattack}, FedAttack \cite{wu2022fedattack}, PipAttack \cite{PipAttack-zhang2022pipattack}, UA-FedRec \cite{UA-FedRec-yi2023ua}, ClusterAttack \cite{ClusterAttack-yu2023untargeted}, IMIA \cite{yuan2023interaction}, Zhang et al. \cite{zhang2023comprehensive}, FRecAttack$^2$ \cite{FRecAttack2-hao2024not}, StairClimbing \cite{StairClimbing-hao2024eyes}, UC-FedRec \cite{UC-FedRec-hu2024user}, CIRDP \cite{CIRDP-liu2024defending}, PoisonFRS \cite{PoisonFRS-yin2024poisoning}, PIECK \cite{PIECK-zhang2024preventing}, PAMN \cite{PAMN-zheng2024poisoning}, HidAttack \cite{HiAttack-ali2024hidattack}, ConDA \cite{ConDA-liang2025defending}, Aegis \cite{Aegis-wu2025aegis}, HMTA \cite{HMTA} & 
Clients train the recommendation model using local data and send the model parameters and the item embeddings containing the sampled pseudo-
items to the server for aggregation. The server is responsible for aggregating these
parameters and selecting the clients to be sent down. Statistical heterogeneity in the system due to non-independent homogeneous distribution of FL client data.& 
\shortstack[l]{\small $\bullet$ Ensuring interaction privacy.\\ \small $\bullet$ Handling model inconsistency. \\ \small $\bullet$ Facilitating client participation. \\\small $\bullet$ High-order interaction mending. \\ \small $\bullet$ Attack and defense.}\\
\hline

Cross-domain FedRec & FedCDR \cite{meihan2022fedcdr}, FedCORE \cite{fedcore}, PPGenCDR \cite{ppgencdr}, FedGCDR \cite{fedgcdr}, PriCDSR \cite{PriCDSR}, FedCT \cite{liu2021fedct}, FPPDM \cite{PP-FedCDR-liu2023federated}, PriCDR \cite{HeteroCDR}, DPSMRec \cite{DPSMRec-liu2023differentially}, FL-MV-DSSM \cite{FL-MV-DSSM}, ReFer \cite{ReFer-li2024refer}, FedDCSR \cite{FedDSCR-zhang2024feddcsr}, FedHCDR \cite{FedHCDR-zhang2024fedhcdr}, P2FCDR \cite{WinWin-chen2023win}, FedCSR \cite{zheng2025fedcsr}, PPCDR \cite{PPCDR-tian2024privacy}, PFCR \cite{PFCR-guo2024prompt} & In most studies, the source of statistical heterogeneity in the system is the label drift carried by clients from different domains. In
addition to aggregating and sharing model parameters, the server is responsible for mitigating this type of data heterogeneity and aligning the representations. &

 \shortstack[l]{\small $\bullet$ User-level relevance.\\ \small $\bullet$ Item-level relevance.\\ \small $\bullet$ Content-level relevance.}\\ \hline

    Multi-modal FedRec &
    AMMFRS \cite{AMMFRS-feng2024recommendation}, FedMMR \cite{FedMMR-li2024towards}, P2M2-CDR \cite{P2M2-CDR:Wang2024APF}, FedMR \cite{FedMR-li2024personalized} &
    The key issue in multi-modal FedRec is to
align the modal and collaborative (ID) features
of items. Typically, the server is responsible
for extracting the modal features, and the
alignment process is chosen to be performed
on the server side or the client side based on
the downstream tasks and client types.
& \shortstack[l]{\small $\bullet$ Aligning modals on the server. \\ \small $\bullet$ Aligning modals on clients.} \\
\hline
    LLM-based FedRec & FedPA \cite{FedPA-zhang2024federated}, PPLR \cite{PPLR-zhao2024llm}, FELLAS \cite{yuan2024fellas}, MRFF \cite{MRRF} & The core of LLM-based FedRec is to perform domain-specific adaptation of the foundation
model. In this scenario, the server is
responsible for aggregating and sharing the
specific parameters, including intrinsic parameters built in the LLM and extrinsic adapters external to the LLM, and selecting the
specific fine-tuning approach in the clients
based on downstream tasks and client types.
& \shortstack[l]{\small $\bullet$ Fine-tuning partial parameters of \\ ~~~~the LLM. \\ \small $\bullet$ Fine-tuning external parameters.} \\
    \hline

\end{tabular}%
}
\label{tab:All_Methods}
\end{table*}

\subsection{Motivation and Contribution}

Since the first FedRec method FCF \cite{FCF-ammad2019federated} has been proposed, related research in this field progressed rapidly.
We summarize the major research efforts since 2019 through a systematic review and present the statistical information in Fig.~\ref{fig:statistic}.
It is evident that an increasing number of studies are focusing on this field, but they are limited to basic scenarios, especially collaborative FedRec.
There is still little research on integrating more contextual information to improve FedRec performance, and there is a lack of exploration of more practical scenarios, severely limiting the deployment of FedRec.

Previous work reviews the FedRec research from the perspective of building typical FL systems, providing rich theoretical insights of extending RS to the FL framework.
For instance, Yang et al. \cite{yang_survey} comprehensively review FedRec across three typical FL levels: horizontal federated systems, vertical federated systems, and federated transfer systems.
Sun et al. \cite{sun_survey} discuss how to address various issues caused by the FL architecture in FedRec, including statistical heterogeneity, communication cost, and privacy protection, which are the most fundamental issues in FL research.
Wang et al. \cite{wang_survey} focus on protecting user privacy and provide an in-depth study of horizontal FedRec from the perspectives of model design, privacy enhancement and system implementation.
Li et al. \cite{li_survey} conduct a comprehensive review of FedRec with a foundation model. They focus on integrating the foundation model to address the limitations of FedRec, and discuss potential future research directions.
Yang et al. \cite{yang2025survey} focus on security and efficiency challenges under cross-user FedRec settings and review related research from four perspectives: privacy, security, accuracy, and efficiency.

However, these reviews treat FedRec as a machine learning task, considering how to perform overall transfer from the perspective of FL scholars. 
Due to the neglect of the unique characteristics and practical challenges of specific recommendation scenarios, their utility drops.
For example, clients in a cross-domain FedRec usually refer to platforms or enterprises holding completely different data, such as e-commerce apps and social media.
There exists a natural statistical heterogeneity and label drift between the data, which is not caused by the FL architecture.
Therefore, it is not sufficient to consider how to mitigate the statistical heterogeneity of this scenario only from the perspective of designing FL systems; one should also consider solving real-world problems in the scenario to encourage deployment.

To bridge this gap, we comprehensively review the state-of-the-art FedRec research from the perspective of recommendation researchers and practitioners. 
Based on the introduction of contexts, we categorize the existing work into four scenarios in Table ~\ref{tab:All_Methods}: collaborative FedRec, cross-domain FedRec, multi-modal FedRec and LLM-based FedRec.
We establish a clear link between these key recommendation scenarios and the FL architecture, systematically discussing and explaining various aspects of them, including approaches, practical challenges and potential opportunities.
Compared to existing reviews, we comprehensively examine the FedRec field through a more practical perspective and provide insights on encouraging realistic deployments.

The main contributions of this review are as follows:
\begin{itemize}
\item \textbf{Explore from a practical perspective}.
This paper thoroughly discusses the coupling between RS and FL. It reviews existing FedRec research from the perspective of building a real-world privacy-protected RS, which significantly differs from existing reviews.
We aim to comprehensively review FedRec's development and provide guidance for real employment.

\item \textbf{Cover multiple fine-grained scenarios}.
We establish clear connections between RS and FL in fine-grained scenarios, including  collaborative FedRec, cross-domain FedRec, multi-modal FedRec and LLM-based FedRec.
This review explores Fedrec's application potential and challenges in these key scenarios, providing valuable insights to promote its effective deployment.

\item \textbf{Concise and comprehensive review}.
This paper comprehensively reviews the state-of-the-art FedRec research in key scenarios, explaining the concepts and techniques involved clearly and concisely.
We organize the existing work according to the characteristics and processes of the scenarios and make visualizations to ensure that both researchers and practitioners can benefit from this review.

\item \textbf{Detailed discussion of key challenges}. 
This review thoroughly discusses the open issues in deploying FedRec across various scenarios and highlights key areas that warrant further investigation.
We carefully explore the links between key FedRec challenges and broader FL topics to gain valuable insights into potential solutions.
Taking the broader view, we further examine existing gaps within the entire FedRec field and propose future directions to promote its development.

\end{itemize}

\section{Procedure of FedRec}
The deployment of recommender systems requires the collaboration of both servers and clients either in CenRec or FedRec. Yet the focus varies across different paradigms. In CenRec, the client primarily serves as a data producer device, while the server is the focal point for model training. To preserve user privacy, FedRec offloads model training to clients, thereby eliminating the server’s access to raw user data. Consequently, while key distinctions exist, FedRec shares certain procedural similarities with CenRec. The following two sections progressively detail their respective workflows.
\subsection{Procedure of CenRec}

Formally, there are a user set $\mathcal{U}$ and an item set $\mathcal{I}$. The basic training data in recommender system is the interaction records between users and items, denoted as $\mathcal{D}$. From the overview, CenRec collects data to form an interaction matrix $\mathbf{R}\in \mathbb{R}^{|\mathcal{U}|\times|\mathcal{I}|}$, where each row represents the interaction record of one individual user. From the specific perspective, the interaction dataset of one user, denoted as $\mathcal{D}^u$, consists of triple tuples as $(u,m,r_{um})$, where $u \in [|\mathcal{U}|]$ and $m \in [|\mathcal{I}|]$. In implicit feedback scenarios \cite{he2016fast}, the ground truth $r_{um}$ represents the behavior (\textit{e.g.} click, view or buy) between user $u$ and item $m$, thus $r_{um} \in [0,1]$. Alternatively, in explicit feedback scenarios \cite{zhang2014explicit}, $r_{um}$ refers to the actual rating score given by user $u$ to item $m$. The recommendation model $\theta$ usually consists of three main modules to capture the collaborative information: the user embedding module $\mathcal{P}$, the item embedding module $\mathcal{Q}$ and the interaction function $\mathcal{F}$, \textit{i.e.},~$\theta=\{\theta_{\mathcal{P}},\theta_{\mathcal{Q}},\theta_{\mathcal{F}}\}$. The recommendation task for mining potential items for users is formalized as follows:
\begin{align}
    \label{eq:user_embd}
    \mathbf{p}^u &=\mathcal{P}(\theta_{\mathcal{P}};u),\\
    \label{eq:item_embd}
    \mathbf{q}^m &=\mathcal{Q}(\theta_{\mathcal{Q}};m),\\
    \label{eq:rec_inter}
    \hat{r}_{um} &= \mathcal{F}(\theta_{\mathcal{F}};({\mathbf{p}^u,\mathbf{q}^m})),
\end{align}
where $\mathbf{p}^u \in \mathbb{R}^d$ and $\mathbf{q}^m\in\mathbb{R}^d$ represent the embedding vector of user $u$ and item $m$ with $d$ dimensions, respectively. We denote the loss function of the recommendation task as ${L}$, and the objective is to solve:
\begin{align}
    \label{eq:rec_ob}
    \mathop{\arg\min}\limits_{\theta}~{L}(\theta) = \mathop{\arg\min}\limits_{\theta}~ \mathbb{E}_{(u,m,r_{um})\backsim\mathcal{D}}[l(\theta;(u,m,r_{um}))].
\end{align}
Thus, the back propagation of the recommender system is formalized as follows:
\begin{align}
    \label{eq:rec_loss}
    \theta = \theta - \eta~\nabla\theta,
\end{align}
where $\nabla\theta = \frac{\partial~{L}(\theta)}{\partial~\theta}$ represents gradients of the recommendation model, and $\eta$ is the learning rate. CenRec conducts model training of $\theta$ on the server, whereas clients serve as nodes responsible for data uploading and service receiving.

\begin{algorithm}[!t]
\caption{Federated Recommendation System}\label{alg:FedRec}
\SetAlgoLined
\DontPrintSemicolon
\KwIn{
    The sampling number of clients $S$, total communication rounds $T$, and the total local training epochs $E$
}
\Comment{\color{blue}On the server}
Initialize global parameter component $G_\theta$ \;
\For{communication round $t = 1$ \KwTo $T$}{    
    Sample $S$ clients $\mathcal{N}_S \subseteq \mathcal{N}$ \;
    Distribute $G_\theta$ to clients in $\mathcal{N}_S$\;
    \Comment{\color{blue}Global model broadcast}
    \For{each client $i \in \mathcal{N}_S$ \textbf{in parallel}}{
        $\Delta G_\theta^{(i)} \gets \text{ClientUpdate}(i, G_\theta)$ \;
        \Comment{\color{blue}Parallel local training}
    }
    $G_\theta \gets \text{Aggregate}\left(\{\Delta G_\theta^{(i)}\}_{i\in\mathcal{N}_S}\right)$ \;
    \Comment{\color{blue}Model aggregation, typically via FedAVG with Eq.~\eqref{eq:FedAVG}}
}
\Comment{\color{blue}On clients}
\SetKwProg{Fn}{Function}{}{}
\Fn{ClientUpdate($i, G_\theta$)}{
    \If{client $i$ debuts in training}{
        Initialize private parameters $V_\theta^i$ \;
    }
    Construct full model $\theta^i = [G_\theta; V_\theta^i]$ \;
    \For{local epoch $e = 1$ \KwTo $E$}{
        $\theta^i \gets \theta^i - \eta\nabla\mathcal{L}(\theta^i;\mathcal{D}^i)$ \;
        \Comment{\color{blue}Update via Eq.~\eqref{eq:rec_loss}}
    }
    \eIf{parameter transmission}{
        \Return $G_\theta^i$ \;
        \Comment{\color{blue}Upload model parameters}
    }{
        \Return $\nabla G_\theta^i$ \;
        \Comment{\color{blue}Upload model gradients}
    }
}
\end{algorithm}

\subsection{Procedure of FedRec}
FedRec decentralizes the training process of the recommendation model $\theta$ by delegating it to service-receiving clients. The server connects data silos across clients through model aggregation, thereby eliminating the force on clients to share their private data.
Assuming there are $N$ clients attending recommendation services, the client set is denoted as $\mathcal{N}$, where the $i$-th client has a local dataset $\mathcal{D}^i$. The objective of FedRec is to solve:
\begin{equation}
    \label{eq:FLobj}
    \mathop{\arg\min}\limits_{\theta}~\mathcal{L}(\theta) = \sum_{i=1}^N\frac{|\mathcal{D}^i|}{|\mathcal{D}|}{L}^i(\theta),
\end{equation}
where ${L}^i(\theta)=\mathbb{E}_{(u,m,r_{um})\backsim\mathcal{D}^i}[l^i(\theta;(u,m,r_{um})]$, and $(u,m,r_{um})$ represents the interaction tuple produced on the individual client. Consequently, FedRec processes $\mathcal{D}\triangleq \bigcup_{i\in \mathcal{N}}\mathcal{D}^i$, which means the universal dataset $\mathcal{D}$ consists of local datasets on each client. In FedRec, the client may represent either the individual user or the recommendation platform/corporation, depending on different levels of protection requirements. In the cross-user scenario \cite{yang2025survey}, each client manages the interaction data of a single user, where $\mathcal{D}^i\equiv\mathcal{D}^u$ and $|\mathcal{N}|=|\mathcal{U}|$. Conversely, in the cross-organization scenario, clients admit interactions of users within one platform or enterprise, where data remains shareable in the same organization yet isolated across organizational boundaries. Thus, this scenario features $|\mathcal{N}|\ll|\mathcal{U}|$.

Notably, \textbf{FedRec does not collaborate the full parameter among clients and the server}, since some parameters are closely tied to individual attributes, and sharing them will compose user privacy. In fact, FedRec categorizes parameters of the recommendation model $\theta$ into global component $G_\theta$ and private component $V_\theta$, \textit{i.e.},~$\theta=\{G_\theta, V_\theta\}$. Typically, the private model component $V_\theta$ contains the user embedding module $\mathcal{P}$, which is preserved locally. Yet the item embedding module $\mathcal{Q}$ and the interaction function $\mathcal{F}$ belong to the global model $G_\theta$, subject to aggregation and sharing among clients and the server.

During each communication round, the server select a subset $\mathcal{N}_S$ with $S$ clients to train the model with their local dataset. After local training, clients package local updates, \textit{e.g.} parameters $G_\theta$ or gradients $\nabla G_\theta$, of the global component and upload them. Then the server aggregates local updates to obtain a new $G_\theta$ for the next communication round. This aggregation procedure commonly integrates the classic algorithm, FedAVG \cite{FedAVG-mcmahan2017communication}, which is defined in the formula as:
\begin{equation}
        \label{eq:FedAVG}
        G_\theta =
        \begin{cases}
        G_\theta - \eta\sum_{i\in\mathcal{N}_S}p^i \nabla G_\theta^i,  & \text{when transmit $\nabla G_\theta$}; \\[2ex]
        \sum_{i\in\mathcal{N}_S}p^i G_\theta^i, & \text{when transmit $G_\theta$,}
        \end{cases}
\end{equation}
where $p^i=\frac{|\mathcal{D}^i|}{\sum_{s\in\mathcal{N}_S}|\mathcal{D}^s|}$.
The detailed training procedure in the FedRec paradigm is illustrated in Algorithm \ref{alg:FedRec}.

\section{Collaborative FedRec}
\begin{figure*}
	\centering
	\includegraphics[width=1\linewidth]{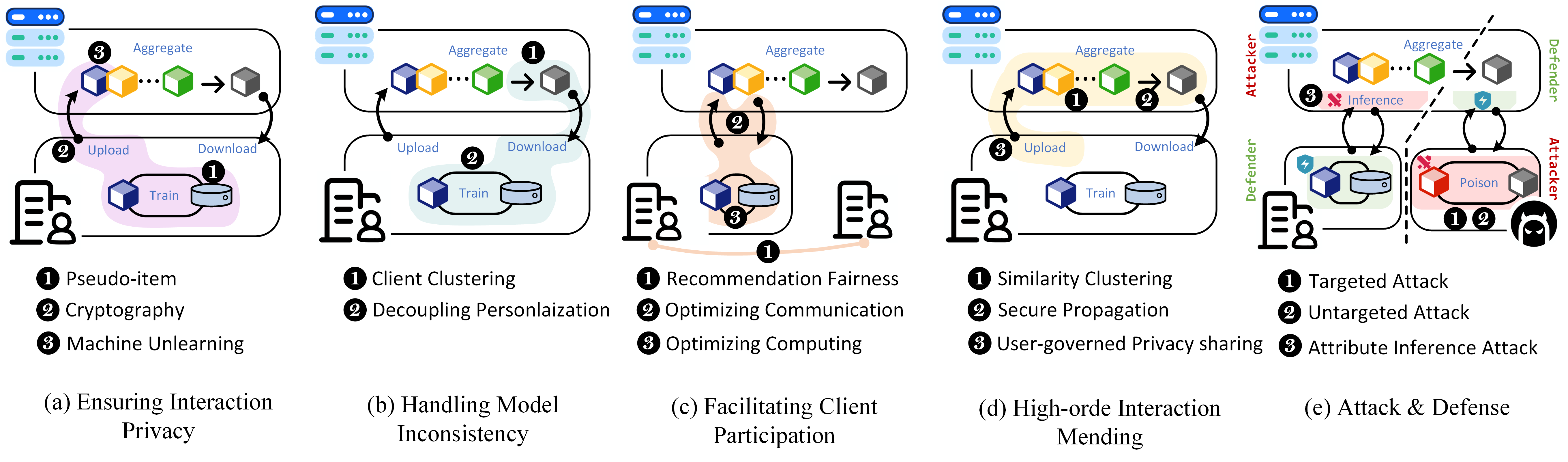}
	\caption{Collaborative Federated Recommendation.}
	\label{fig:CollaborativeRS}
\end{figure*} 

Collaborative RS faces two primary challenges: effectively learning user and item representations and developing accurate interaction modeling methods. Research in collaborative CenRec can thus be broadly categorized into two branches: 
 1) \textbf{Representation Learning} involves learning embedding representations for each user and item based on their interaction patterns. The core challenge lies in deriving accurate embeddings from typically sparse interaction data. Traditional matrix factorization methods \cite{BPR-rendle2012bpr,MF-koren2009matrix,FAWMFchen2020fast} have proven efficiency. To further improve performance, some approaches incorporate attention mechanisms to weight user interaction histories \cite{NAIS-he2018nais, ACL-chen2017attentive}. Autoencoder-based models, leveraging the principle of input reconstruction, have also gained prominence \cite{CDAE-wu2016collaborative,Mult-VAE-shenbin2020recvae,SW-DAE:khawar2020learning}. While these methods focus on individual user interaction data, the advent of graph neural networks (GNNs) \cite{DGCF:wang2020disentangled,LightGCN-he2020lightgcn,DHCF-ji2020dual,LR-GCCF:chen2020revisiting} has enabled the propagation of higher-order interaction information by effectively connecting first-order interactions across users.
 2) \textbf{Interaction Modeling} aims to estimate the user preference for different items based on user and item embeddings. Traditionally, the inner product of user and item embeddings has been employed for this purpose. Despite its simplicity and effectiveness, the inner product suffers from two limitations: it neglects user-user and item-item relationships, and it struggles to capture complex, non-linear user-item interactions. To address the former, distance-based metrics have been proposed as alternative interaction functions \cite{CML-hsieh2017collaborative, TransRec:he2017translation, LRML:tay2018latent, HyperML:vinh2020hyperml}. To address the latter, more recent work has explored leveraging neural network architectures, such as multi-layer perceptrons (MLPs) \cite{NCF-he2017neural}, convolutional neural networks (CNNs) \cite{ONCF:he2018outer}, and autoencoders \cite{CDAE-wu2016collaborative,Mult-VAE-shenbin2020recvae}, to capture the intricacies of user-item interaction.

Due to the efficient and widespread deployment of collaborative recommender systems, research on collaborative FedRec has emerged as a primary focus within FedRec. This setting typically treats each individual user as a separate client, reflecting the emphasis on highly personalized services -- a characteristic aligned with CenRec. 
We categorize the concerns of existing collaborative FedRec research across four dimensions: data, model, participant, and safeguard. We further offer a holistic perspective and illustrate the most crucial issues on current collaborative FedRec research from the following five problems: ensuring interaction privacy, handling model inconsistency, facilitating client participation, high-order interaction mending, and attack \& defense. Fig.~\ref{fig:CollaborativeRS} briefly shows the corresponding challenges.

\subsection{Ensuring Interaction Privacy} \label{sec:privacy} 

Model gradients or parameters uploaded by clients can still inadvertently reveal user behaviors. This risk stems from the fact that users typically interact with only a subset of items, and partial updates to item embeddings, which are integral to model updates, can expose patterns in user behavior. Additionally, in explicit recommendation scenarios, the magnitude of gradient updates—driven by high or low ratings—can further expose user preferences \cite{FedRec-lin2020fedrec, FedRec++-liang2021fedrec++}. Consequently, ensuring interaction privacy from model or gradient-based inference attacks is a critical concern in Collaborative FedRec.

The related research can be categorized based on the technique as follows:
\textbf{(1) Pseudo-item}. A common approach to obfuscating user behavior involves randomly sampling uninteracted items on the client-side and transmitting these to confuse the server as pseudo-interactions. Examples of this technique include FedRec \cite{FedRec-lin2020fedrec} and FR-FMSS \cite{FR-FMSS-lin2021fr}. However, they introduce noise during server-side aggregation, potentially degrading recommendation performance. FedRec++ \cite{FedRec++-liang2021fedrec++} attempts to mitigate this by exploring denoising strategies within the client pool to facilitate noise-free interaction masking.
\textbf{(2) Cryptography}. Like other FL topics, collaborative FedRec systems leverage cryptographic techniques to enhance protection against inference attacks targeting model gradients or parameters. Commonly employed methods include homomorphic encryption \cite{FedMF-chai2020secure}, hash mapping \cite{LightFR-zhang2023lightfr}, and differential privacy \cite{SP-FCF-StrongPrivacyCF-minto2021stronger}. These techniques conceal private information from the server by mapping raw data into an encrypted space. 
\textbf{(3) Machine unlearning}. This technique aims to remove the influence of specific data from a trained model retroactively, as if the data had never been used for training. FRU \cite{FRU-yuan2023federated} applies this concept to address privacy requirement, allowing users to retract their interactions to the federated model. This is achieved by rolling back and calibrating historical updates, which are then used to expedite the reconstruction of FedRec.

\subsection{Handling Model Inconsistency} \label{sec:model} 

User interaction patterns are shaped by dual factors: collective-level popularity trends and individual preferences, possessing significantly non-independent and identically (non-iid) distribution in FedRec. Each client optimizes its own model to fit local preference distributions, which introduces critical challenges during aggregation. Divergent gradient updates across clients substantially hinder global model convergence. The global model struggles to preserve node-specific preference features during parameter averaging, resulting in feature shifts and systematic representation degeneration.

To address this concern, existing research focuses on balancing personalization and generalization in collaborative FedRec, primarily through two approaches: 
\textbf{(1) Client Clustering}. 
Clustering clients with similar preferences enables knowledge transfer through cluster-level model aggregation, while maintaining a degree of personalization. For instance, FedFast \cite{FedFast-muhammadFedFastGoingAverage2020a}, PerFedRec \cite{PerFedRec-luo2022personalized}, PerFedRec++ \cite{PerFedRec++-luo2024perfedrec++}, and PrefFedPOI \cite{PrefFedRec-zhang2023fine} all propose cluster-based aggregation methods to address individual preferences. Additionally, GPFedRec \cite{GPFedRec-zhang2024gpfedrec} introduces a graph-based aggregation strategy on the server, utilizing item embedding clustering to generate personalized models for each user.
\textbf{(2) Decoupling Personalization}. This approach involves introducing personalized models on the client side that do not participate in the aggregation process, allowing for further personalization of the recommendation model. Examples include local fine-tuning strategies in PFedRec \cite{PFedRec-zhang2023dual} and PrefFedPOI \cite{PrefFedRec-zhang2023fine}, as well as the additive personalized local item embedding table in FedRAP \cite{FedRAP-lifederated}.

\subsection{Facilitating Client Participation} \label{Sec:facilitate}

Since edge devices commonly serve as clients in collaborative FedRec, significant fairness concerns arise due to disparities in computational power, communication capabilities, and data resources. FL often prioritizes clients with abundant interaction data, robust hardware, or stable network conditions during training. This bias leads to imbalanced participation: resource-rich clients bear excessive computational and communication burdens, while those with limited capabilities risk exclusion or delayed synchronization with the global model. Such inequities perpetuate performance gaps and undermine recommendation fairness, diversity, and system sustainability.

To foster inclusive and efficient participation, recent research focuses on three critical dimensions:
\textbf{(1) Ensuring Recommendation Fairness}. In addition to federated fairness issues concerning client participation, traditional fairness concerns in recommendation scenarios (\textit{e.g.} age, gender) remain relevant. Privacy restrictions make it difficult for the server to access clients’ actual attributes, further complicating ensuring fairness in recommendations. F$^2$PGNN \cite{F2PGNN-agrawal2024no} encodes fairness constraints as a regularization term and leverages demographic group statistics collected from clients to enhance group fairness. IFedRec \cite{IFedRec-zhang2024federated} focuses on users who interact with long-tail items by leveraging a meta-attribute network to transfer patterns from warm to cold items.
\textbf{(2) Reduction in Communication Costs}. Learning item relationships requires coordination among multiple clients. The limited communication resources of users' private devices highlight a critical issue in efficiently transmitting models. LightFR \cite{LightFR-zhang2023lightfr} enhances inference and communication efficiency by generating high-quality binary codes using learning-to-hash techniques in federated settings. MetaMF \cite{MetaMF-lin2020meta} and ReFRS \cite{ReFRS-imran2023refrs} reduce communication costs by generating item embeddings based on client preferences. PO-FCF \cite{PO-FCF-khan2021payload} employs reinforcement learning to select subsets of items with positive feedback for FL training, adaptively minimizing communication overhead. CoLR \cite{CoLR-nguyen2024towards} leverages low-rank approximations of item embeddings to lower communication costs without adding computational burden and is compatible with techniques like homomorphic encryption. PTF-FedRec \cite{PTF-FedRec-yuan2024hide} collaboratively exchange knowledge by sharing model predictions instead of high-dimensional parameters within a privacy-sampling strategy.
\textbf{(3) Reduction in Computation Costs}. For client devices, training large models on local datasets is often unfeasible, severely limiting the participation of lighter clients in attaining quality recommendation. RF$^2$ \cite{RF2-maeng2022towards} implements tier-aware gradients pruning to adapt the model to local devices. Efficient-FedRec \cite{Efficient-FedRe-yi2021efficient} decomposes the news recommendation model into a large one maintained on the server and a lightweight user model. Recently, HeteFedRec \cite{HeteFedRec-yuan2024hetefedrec} introduces a framework enabling clients to adjust model size based on local interaction volume, using relation-based ensemble knowledge distillation to share information across heterogeneous models.

\subsection{High-order Interaction Mending} \label{sec:inter-mending}

Utilizing graph neural networks (GNN) to capture high-order interaction patterns on the user-item interation is widely adopted in both business application and academic research \cite{LightGCN-he2020lightgcn}. However, the distributed storage of data and the restriction on direct access in collaborative FedRec cut off user-user links, preventing different clients from perceiving intricate interaction patterns of users from other clients. This poses challenges for leveraging graph aggregation methods to mine user interests in collaborative FedRec.

Expanding the local ego-graph to capture higher-order knowledge in a privacy-preserving manner is a unique and critical challenge for Collaborative Federated Recommendation. Researches of mending the high-order interaction progress in three categories:  
\textbf{(1) Similarity Clustering.}
Using encrypted information, recommender systems can achieve privacy-preserving graph expansion on clients. FedPerGNN \cite{FedPerGNN-wu2022federated} introduces a third-party platform to match users with similar interaction behaviors based on homomorphically encrypted item IDs, allowing related user embedding vectors to be delivered to target clients to complete local subgraph expansion. This insight is also adopted and enhanced in SemiDFGL \cite{Semi-qu2023semi} and F$^2$PGNN\cite{F2PGNN-agrawal2024no}.
\textbf{(2) Secure Representation Propagation.} 
Research under this insight allows clients to gain external interaction knowledge without relying on encryption algorithms.
P-GCN \cite{P-GCN-hu2023privacy} uses group-wise concealing and secure aggregation to train a federated recommender system with a global interaction graph. 
FedHGNN \cite{FedHGNN-yan2024federated} performs metapath-based recommendations by leveraging global knowledge, \textit{e.g.},~item types, and perturbed local interactions to help the server reconstruct a global heterogeneous information network.
HiFGL \cite{HiFGL-guo2024hifgl} develops neighbor-agnostic aggregation to cover adjacency information, and then uses Lagrange polynomial functions to securely mask shared node embeddings during information transfer.
VerFedGNN \cite{VerFedGNN-mai2023vertical} transmits the summation of neighbor embeddings through random projection and gradients of global parameters perturbed by a ternary quantization mechanism.
\textbf{(3) User-governed Privacy Sharing.} 
Recently, CDCGNNFed \cite{CDCGNNFed-qu2024towards} introduces a novel framework considering the divergence in privacy protection among users, enabling clients to transmit a selective set of interaction items. The server enhances the user-governed information connectivity by pruning and restoring the existing graph, resulting in a more cohesive interaction graph.

\subsection{Attack and Defense} \label{attack}

FedRec, using a distributed training framework with frequent multi-client communication, inevitably faces attack and defense challenges. Stemming from federated learning, research on poisoning attacks, aiming to promote specific items for economic gain, and untargeted attacks, aiming to degrade the overall recommendation performance, has flourished recently. Additionally, inference attacks targeting tracking user attributes or interaction relationships are gaining attention due to participants' unique characteristics. 

\textbf{(1) Targeted Attack.} Rong et al. \cite{A-ra-rong2022poisoning} first explore attack paradigms A-ra and A-hum in a FedRec, where malicious clients purposefully learn target exposure items based on estimated user vectors to maximize exposure benefits for user groups. FedRecAttack \cite{FedRecAttack-rong2022fedrecattack} further leverages global correlated behaviors (\textit{e.g.}, reviews, follows) to precisely model malicious clients’ feature vectors, enhancing the efficiency of gradient poisoning attacks for boosting specific item exposure. PipAttack \cite{PipAttack-zhang2022pipattack} aligns target item embeddings with popular items and constrains the distance between poisoning gradients and original gradients to evade detection. Hao et al. \cite{StairClimbing-hao2024eyes} propose StairClimbing, constructing user profiles via target-similar items, and introduce CrossEU to analyze benign users’ dual update patterns for identifying malicious gradient updates. PoisonFRS \cite{PoisonFRS-yin2024poisoning} crafts a target model from globally popular items using fake users to shift the model toward target convergence. PIECK \cite{PIECK-zhang2024preventing} identifies popular items via embedding changes during training and employs two strategies—item popularity enhancement and malicious user embedding approximation—to facilitate target exposure. PAMN \cite{PAMN-zheng2024poisoning} implements poisoning attacks in peer-to-peer settings and proposes user-level gradient clipping with sparsified updating for defense. HidAttack \cite{HiAttack-ali2024hidattack} designs a bandit model on attack clusters to dynamically infer optimal poisoning gradients, mitigating Byzantine clients.

\begin{figure*}
	\centering
	\includegraphics[width=0.9\linewidth]{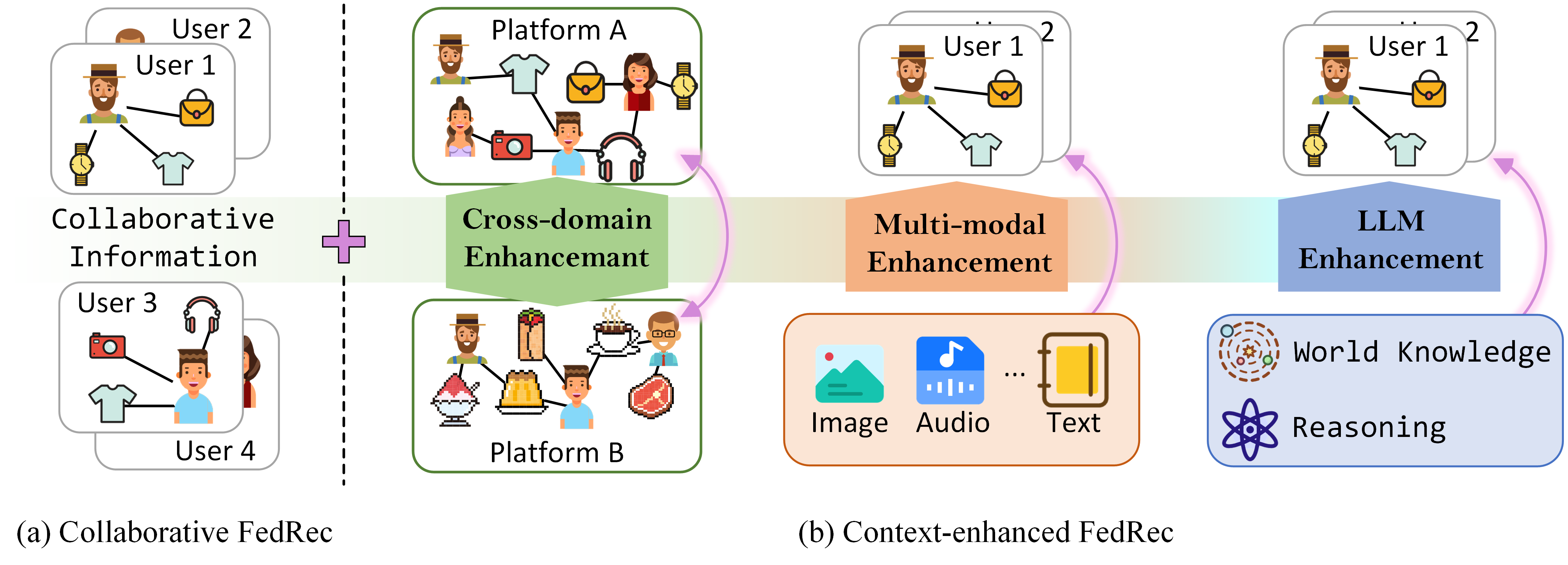}
	\caption{Commonalities and differences of FedRec between the collaborative and context-enhanced scenarios.}
	\label{fig:FCRvsCeFR}
\end{figure*} 

\textbf{(2) Untargeted Attack.} FedAttack \cite{wu2022fedattack} infers user embeddings via profiles and employs globally hard negative sampling items  \cite{fedattack-xuan2020hard}  as preference learning to degrade overall performance. UA-FedRec \cite{UA-FedRec-yi2023ua} learns inverse interaction patterns and fabricates interaction counts to amplify poisoning gradients’ impact during aggregation. ClusterAttack \cite{ClusterAttack-yu2023untargeted} models item embeddings in dense clusters to generate indistinguishable ratings, while proposing UNION—a defense method enforcing uniform embedding distributions via contrastive learning. Hao et al. \cite{FRecAttack2-hao2024not} explore user-item interplay in FRecAttack$^2$ and introduce GuardCQ for malicious interaction detection via contribution quantification. ConDA \cite{ConDA-liang2025defending} computes client-side contribution values, then achieves attack-robust gradient aggregation through decentralized voting and knowledge distillation.

\textbf{(3) Attribute Inference Attack.} Yuan et al. \cite{yuan2023interaction} propose relational inference attacks targeting user interactions, where servers infer actual interacted items by fitting client-uploaded models via shadow recommenders. They introduce a local regularization strategy to constrain global parameter learning and mitigate information leakage \cite{yuan2023interaction}. UC-FedRec \cite{UC-FedRec-hu2024user} addresses the consented attribute sharing issue by applying a sensitive attribute information filter to counter user attribute inference attacks. CIRDP \cite{CIRDP-liu2024defending} incorporates causal inference-augmented offline reinforcement learning and a differentially private representation learning-based defender to resist membership inference attacks. Zhang et al. \cite{zhang2023comprehensive} analyze vulnerability variations in GCN-based FedRec components against attribute inference attacks and introduce adaptive differential privacy strategies for precise defense. Aegis \cite{Aegis-wu2025aegis} employs machine unlearning to eliminate sensitive attributes in user embeddings, coupled with an information-theoretic multi-component loss function to balance privacy protection and recommendation.

\subsection{Challenges} 
The foundational challenges have been preliminarily explored based on existing research in Collaborative Federated Recommendation systems. By analyzing current studies' motivations and technical characteristics and incorporating the user-centered perspective, we identify and summarize the potential challenges inherent in the collaborative FedRec.
\subsubsection{Trade-off between Privacy and Efficiency}
Existing approaches often leverage encryption or obfuscation techniques, such as homomorphic encryption or differential privacy, to avoid privacy leakage from model gradients and parameters.
While highly reliable, these methods impose substantial computational overhead for encryption and decryption, especially for resource-constrained private devices. 
Furthermore, differential privacy and obfuscation strategies, such as pseudo-interactive item sampling, introduce noise into the model, potentially compromising its ability to characterize users and items accurately. Recent efforts have sought to reduce the computational burden of privacy-preserving strategies, proposing efficient encrypted transmission methods based on hash mapping \cite{FRU-yuan2023federated} and various secret-sharing mechanisms \cite{HiFGL-guo2024hifgl, VerFedGNN-mai2023vertical}. However, the corresponding research remains confined to relatively simple model architectures. Developing efficient and architecture-friendly privacy-preserving mechanisms remains a critical and challenging issue for collaborative FedRec.

\subsubsection{Over-personalization}
Personalized FL strategies have proven effective in addressing statistical heterogeneity. Research as PFedRec \cite{PFedRec-zhang2023dual} and FedRAP \cite{FedRAP-lifederated} also demonstrate outstanding performance in collaborative FedRec.
Does this mean that personalized strategies are inherently superior to generalized approaches? Not necessarily. These methods have primarily been validated in relatively simple and constrained interaction scenarios, where collaboration is based solely on user-item interactions. When it comes to more complex contexts, \textit{e.g.} multi-modal recommendation, personalization-centric strategies have not achieved similarly remarkable results \cite{FedMR-li2024personalized}. Collaborative FedRec aims to develop generalized models through collaboration while simultaneously learning personalized components to accommodate diverse user preferences \cite{zhang2023personal}. Striking an effective balance between these two aspects across different scenarios—maximizing their respective strengths—remains a compelling and meaningful research direction in collaborative FedRec.

\section{Context-enhanced FedRec}

In CenRec, item contextual information is widely used to enrich user profiles and mitigate cold-start problems. Thus context-enhanced recommendations prosperous has flourished in industry. Furthermore, since item information is usually considered less privacy-sensitive, incorporating it into FedRec typically does not introduce additional privacy concerns. This makes leveraging item context a promising technique for FedRec.
In this chapter, we review the relevant work and propose potential issues of introducing different types of contexts into FedRec in detail, which is briefly illustrated in Fig.~\ref{fig:FCRvsCeFR}.
Based on the type of context, we categorize current research efforts into three key areas: cross-domain FedRec, multimodal FedRec, and LLM-based FedRec.

\subsection{Cross-domain FedRec}
Due to the existence of common items or content (\textit{e.g.}, tags, categories) across different domains and the fact that user preferences in multiple domains are often relevant, collaboratively using cross-domain information to alleviate data sparsity and cold-start issues has become a promising solution.
Cross-domain recommendation (CDR) aims to align domain representations and improve recommendation accuracy by leveraging relevant information across these domains, both in centralized and federated architectures. 
The benefit of expanding this scenario to the FL framework is intuitive, as data in different domains are held by different platforms or organizations.
Cross-domain FedRec can alleviate the problem of data silos while protecting data from leakage, achieving a win-win situation among organizations.
Based on the type of relevant information across domains, existing work can be divided into three perspectives \cite{CDRsurvey}: user-level relevance, item-level relevance, and content-level relevance.

\begin{figure*}
	\centering
	\includegraphics[width=1\linewidth]{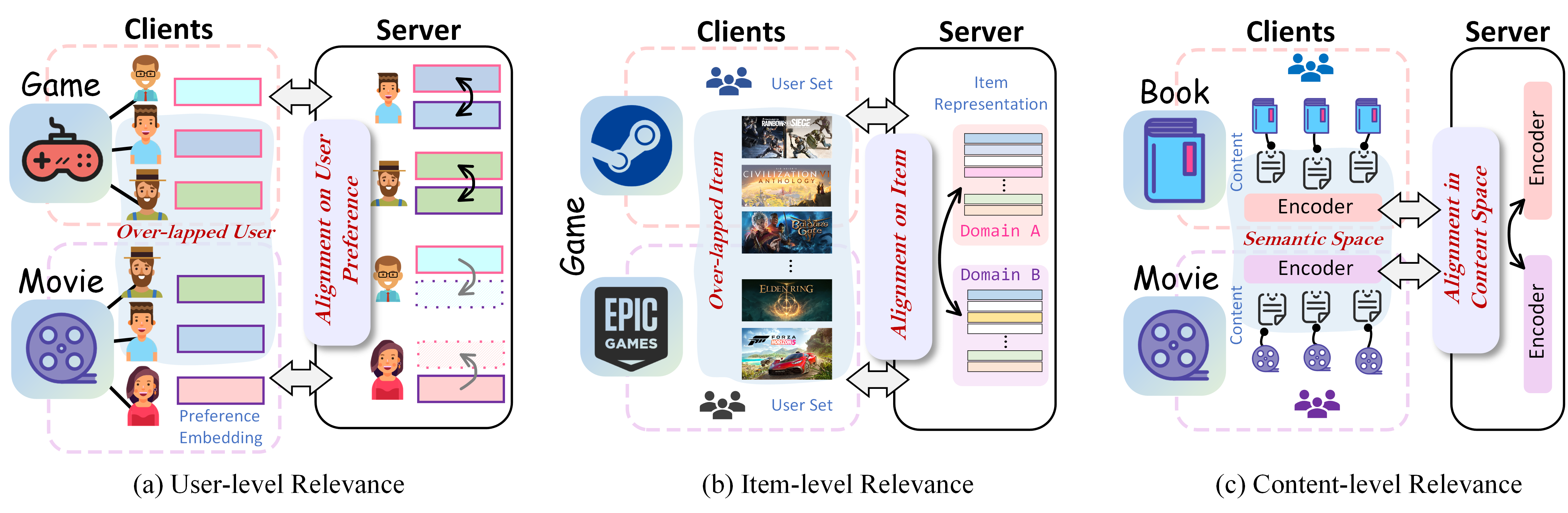}
	\caption{Cross-domain federated recommender systems of three different frameworks.}
	\label{fig:CDFedRec}
\end{figure*} 

\subsubsection{User-level Relevance}
Since user behaviors are scattered across multiple domains and there are correlations between preferences across domains, the most intuitive approach is to align domain representations using overlapping user behaviors and preferences.
Several work \cite{meihan2022fedcdr, fedcore, ppgencdr, fedgcdr, PriCDSR} uses cryptographic methods (\textit{i.e.}, differential privacy and homomorphic encryption) to achieve privacy-preserving cross-domain FedRec.
They encrypt overlapping user behavior and preference features in the source domain (with richer data) and transfer them to the target domain (with fewer data) to improve its performance.
These approaches trade off privacy, data heterogeneity, and communication cost issues within a federated architecture, improving the performance of the target domain.
However, significant statistical heterogeneity in cross-domain FedRec often leads to domain drift, which hinders the direct transfer of knowledge from the source to the target domain.
Consequently, some methods employ indirect transfer and alignment across domains via an intermediate variable \cite{liu2021fedct, PP-FedCDR-liu2023federated, HeteroCDR, DPSMRec-liu2023differentially, FL-MV-DSSM}. 
For example, FedCT \cite{liu2021fedct} learns a shared user representation to achieve unified intra-domain feature learning and efficient inter-domain knowledge transfer.
However, these efforts typically only focus on overlapping users, ignoring the much larger proportion of non-overlapping users in reality.
To this end, ReFer \cite{ReFer-li2024refer} proposes a retrieval-enhanced cross-domain recommender that systematically improves the representation of all users in the FedRec through a generic 'retrieval-and-fusion' framework, achieving a better performance.
 Due to the partial similarity of inter-domain features, some studies decouple features into shared and domain-specific components. 
 FedDCSR \cite{FedDSCR-zhang2024feddcsr} and FedHCDR \cite{FedHCDR-zhang2024fedhcdr} learn the shared and specific perspectives of items to encourage efficient cross-domain alignment and sufficient personalization.
 PPCDR \cite{PPCDR-tian2024privacy} incorporates a graph transfer module in each domain to integrate global and local user preferences.
Besides, P2FCDR \cite{WinWin-chen2023win} learns a bidirectional mapping matrix to achieve mutual enhancement between the source and target domains to enable the source domain to benefit from cross-domain learning.

\subsubsection{Item-level Relevance}
In the global business environment, multinational enterprises often face data compliance barriers across multiple countries—even the subsidiaries of the same organization are unable to share user-related data directly due to localized data privacy regulations.
The key challenge in this scenario lies in reconciling business and item relevance across domains with strictly isolated user groups.
FedCSR \cite{zheng2025fedcsr} addresses these challenges through a dual contrastive learning architecture.
The authors align the global-local model with eliminating distribution bias caused by data silos and dynamically enhance preferences within domain users, providing a promising solution for these issues.

\subsubsection{Content-level Relevance}
Compared to user-level relevant methods, item-level alignment methods have natural privacy-preserving advantages. However, their utilities are limited by the practical constraint of no intersection of cross-domain items.
Recently, some approaches \cite{universal1, universal2, universal3, universal4, universal5, universal6} assume that items with similar semantics have similar descriptions and extract the common semantic representation of multi-domain items through NLP technologies to achieve uniform cross-domain knowledge transfer.
In FedRec, PFCR \cite{PFCR-guo2024prompt} uses multi-domain item descriptions to learn unified item embeddings, with each domain individually constructing user features using a private sequence encoder, building a bridge under the premise of privacy compliance.

\subsubsection{Open Problems and Future Directions}
Existing work migrates CDR to FL frameworks in several ways and demonstrates promising performance.
However, their practicality drops due to the neglect of the following issues:

\textbf{Leakage of overlapping users.} 
Existing work typically assumes that different domains are aware of overlapping users between them, which may cause some practical problems. 
Firstly, knowing the domain of user behaviors can leak their identity information and patterns, violating user privacy protection requirements under regulations such as GDPR.
Secondly, sharing information about overlapping users will increase the risk of data leakage. If data in one domain is leaked, it may affect users' privacy in other domains.
Most importantly, there may be conflicts of interest or competitive relationships between different domains. A domain with overlapping user information may use the data for inference attacks to gain unfair competitive advantages.
The above issues lead to a breakdown in the trust relationship between domains and hinder the growth of FedRec field.

\textbf{Alignment under pseudo-interaction.}
The common practice to protect users' real interaction data is to sample some un-interacted items for each user as noise and exchange the item embedding tables containing noise.
This increases the difficulty of cross-domain alignment, as the system needs to align the semantics of cross-domain items under a large amount of behavioral noise, reducing the practical utility of cross-domain FedRec.
While there are some approaches to cross-domain alignment through item-related information, they do not directly optimize the user's behavioral sequences. Besides, they require careful handling of issues such as anisotropy and uncertainty due to noisy item contents.

\textbf{Cross-domain sequence recommendation.}
In real-world scenarios, user behavior is usually scattered across multiple domains.
The user behavior sequences across domains are essential for CDR, as they help model users' long-term and short-term preferences.
However, related work in cross-domain FedRec prohibits directly sharing client interactions, making it difficult to access and exploit cross-domain sequence information.
Although the authors of FedCSR \cite{zheng2025fedcsr} work on this scenario, they are committed to cross-platform tasks and assume that cross-domain information within the platform is shared.
Therefore, there is an urgent need to explore techniques for obtaining and exploiting information about a user's cross-domain sequential behavior in cross-domain FedRec.

\begin{figure*}[t]
	\centering
	\includegraphics[width=0.95\linewidth]{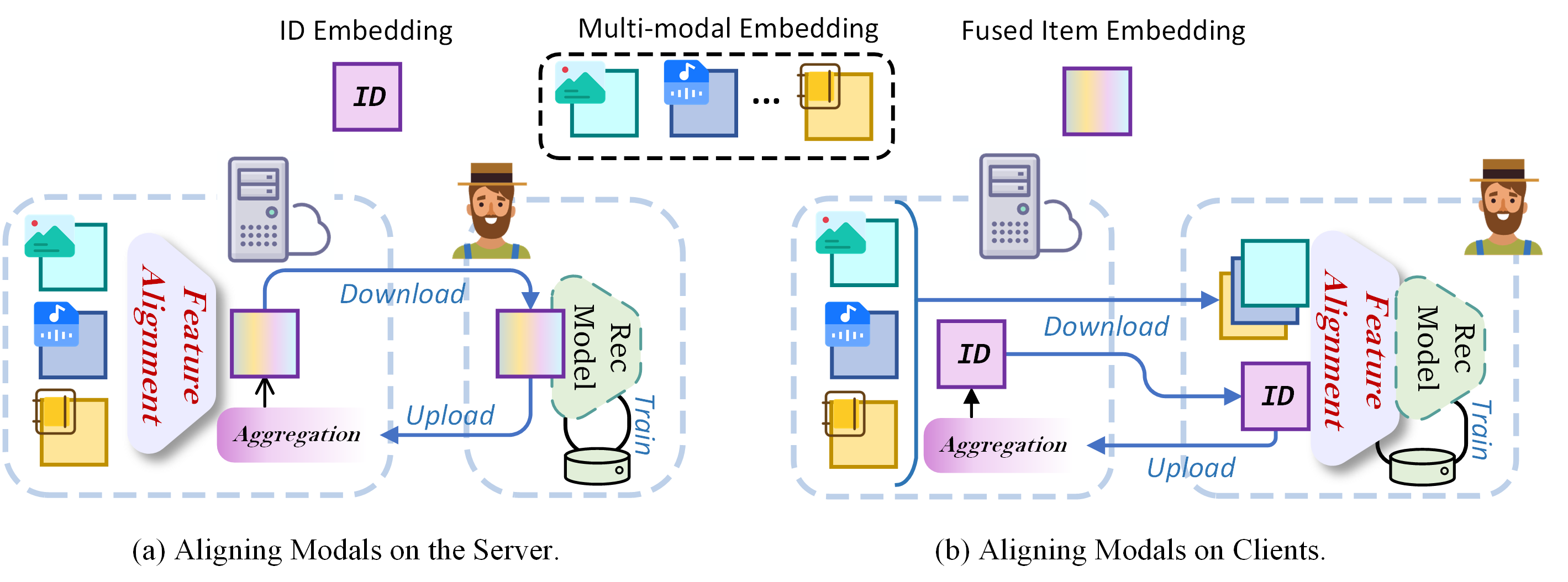}
	\caption{Multi-modal federated recommender systems of two different frameworks.}
	\label{fig:MMFedRec}
\end{figure*} 

\subsection{Multi-modal FedRec}

Multi-modal recommendation (MMRec) systems aim to improve recommendation performance and enhance user experience by involving multiple modalities of data, \textit{i.e.},~audio, images and text \cite{MMRecSurvey-zhou2023comprehensive}. MMRec regards item ID as a foundation information, which captures the characteristics of neighboring items \cite{MMRecID-yuan2023go}, and combines it with auxiliary multi-modal data to enhance item representation \cite{MMRec-zhang2024disentangling}.

MMRec focuses on raw feature representation, feature interaction, and recommendations to make full use of the highly abundant item information \cite{MMRecSurvey-Liu}. \textit{Raw feature representation} aims to exact the raw tabular and multi-modal attributes for following procedures \cite{MMEmbedding-jeong2024demystifying, MMEmbedding2-zhu2022bars}. \textit{Feature interaction} make the system uniform the semantic spaces \cite{MMInteraction-lian2018xdeepfm,DCNv2-wang2021dcn} and align features with user multi-modal preferences \cite{MMRec-zhou2023attention, MMrec-liu2019user}. During the final \textit{Recommendation}, the system needs to uncover potential items under problems like the interaction sparsity \cite{MMRecSparse1-liu2022disentangled, MMRecSparse2-liu2022multi}.
Centralized architectures meet limitations when applied to large-scale, multi-domain data integration \cite{FedMAC}, primarily concerning data privacy, security, and ownership issues in recommendations. Federated learning serves as a privacy-preserving framework that facilitates the integration of diverse multi-modal recommendation scenarios.
The primary research focuses on aligning multi-modal features with the recommendation task under the federated learning paradigm. Thus, existing approaches fall into two categories: server-side feature alignment for robust feature learning and client-side alignment for personalized recommendation modeling.

\subsubsection{Aligning Modals on the Server}
As illustrated in Fig.~\ref{fig:MMFedRec}(a), the server aligns multi-modal embeddings with ID embeddings to integrate the collaboration pattern. The insight comes to derive a unified item representation aligned with the interaction pattern from the original feature representations of various item modalities.   
Therefore, this paradigm does not incur additional computational overhead from incorporating of rich modal features, making it more suitable for deployment on edge devices.
AMMFRS \cite{AMMFRS-feng2024recommendation} constructs a personalized-generalized architecture by introducing a penalty term and employs attention mechanisms to aggregate information across modalities. Considering the inexpensive storage and computational resource consumption of multi-modal encoders, FedMMR \cite{FedMMR-li2024towards} delegates multi-modal learning to a powerful server, using contrastive learning to align auxiliary modalities such as visual or textual features with ID representations, training separate encoders for each modality. 

\subsubsection{Aligning Modals on Clients}
As illustrated in Fig.~\ref{fig:MMFedRec}(b), clients train a fusion module locally using interaction data to obtain personalized multi-modal item representations.
Emphasizing the role of clients in leveraging multi-modal embeddings is based on the inconsistency of ID embeddings caused by the heterogeneity of local interactions. Thus, designating clients as the primary agents for feature alignment inherits the collaborative FedRec strategy for handling model inconsistency, aiming to mitigate the non-IID issue through additional personalized models.
P2M2-CDR \cite{P2M2-CDR:Wang2024APF} utilizes MLPs to decouple domain-common and domain-specific embeddings for multi-modal data in user-concerned federated cross-domain recommendation. Owing to the limited resources in the cross-device scenario, FedMR \cite{FedMR-li2024personalized} incorporates a foundational model on the server to extract item features such as images or text and then sends these representations to the client. 

\subsubsection{Open Problems and Future Directions}
Existing research on federated multi-modal recommender systems remains limited. It strongly emphasizes exploration and migration but lacks in-depth investigation and overlooks critical issues. To offer a more valuable perspective for future research, we combine insights in federated multi-modal learning with multi-modal FedRec research and summarize the challenges of multi-modal FedRec tasks.
    
\textbf{Heterogeneous Alignment.} 
Research on centralized multi-modal recommender systems has provided alignment strategies for distinct and sparse semantic space modalities \cite{ fusion-yu2023multi,fusion-zhou2023attention}, because the recommendation performance and generalization of MMRec will be significantly improved if the unique and common characteristics can be distinguished \cite{MMRecSurvey-Liu}. However, existing federated studies still lack in-depth discussion on modality alignment scenarios addressing inherent heterogeneity caused by both unique or common semantic modal information and non-iid interactions. The sparsity of local interactions and heterogeneity among clients present challenges for feature fusion in multi-modal recommendations. Collaboratively aligning practical multi-modal information with interaction patterns in this context attaches a challenging yet valuable research direction.

\textbf{Incongruent Multi-modal Scenario.}
Given the potential of multi-modal FedRec to bridge data silos without direct data transfer, they hold significant promise for cross-platform applications. They primarily face the challenge of incongruent multi-modalities across domains, manifesting as \textit{Exclusive Modality} and \textit{Overlapping Modality} \cite{FedMM-che2023multimodal, FedMM-pan2024survey}. Clients contribute entirely different primary modalities in the former, \textit{e.g.} music-book cross-domain recommendations. In the latter, there is partial overlap in data modalities across clients, \textit{e.g.} blog-book recommendations. While this issue has been extensively studied in federated multi-modal learning, \textit{i.e.} exclusivity \cite{ME-dai2024federated, ME-yang2024cross} and overlap \cite{MO-chen2022fedmsplit, MO-feng2023fedmultimodal, MO-guo2023pfedprompt, MO-zong2021fedcmr}, it remains underexplored in the context of recommendation tasks.

\begin{figure*}[t]
	\centering
	\includegraphics[width=0.8\linewidth]{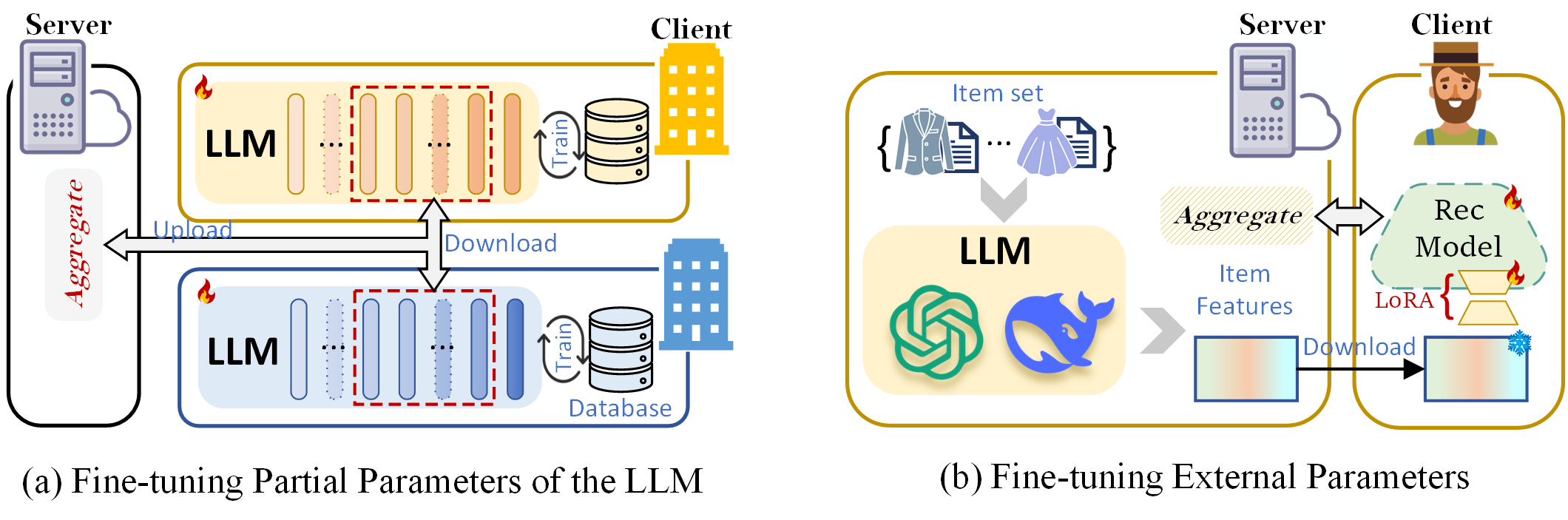}
	\caption{Different federated LLM-based recommendation architectures.}
	\label{fig:fedllm4rec_emb}
\end{figure*}

\subsection{LLM-based FedRec}

With the emergence of large language models (LLM) in natural language processing \cite{LLM-brown2020language}, computer vision \cite{LLM-zhou2020unified}, and molecule discovery \cite{LLM-li2024empowering}, there are increasing researchers exploring the way to enhance RS. Trained via self-supervision on massive datasets, LLM shows promising performance in learning universal representations \cite{LLM-wu2024survey} and offer potential improvements across various recommender system aspects through techniques like fine-tuning \cite{Tuning1-wu2021empowering, Tuning2-qiu2021u} and prompt tuning \cite{Prompt2-yang2022improving}. LLM's strength lies in extracting high-quality textual feature representations and leveraging their encoded external knowledge \cite{LLMSurvey-liu2023pre}.
Specifically, the application of the LLM in RS mainly includes:
1) \textbf{Representation Enhancement.} LLM offers powerful natural language processing capabilities, enabling their application in RS to enrich user and item representations with external knowledge \cite{P5-geng2022recommendation, IDLLM-hua2023index, IDLLM-rajput2023recommender, textLLM-fan2023zero, textLLM-yuan2023go}.
This greatly alleviates RS's cold start problem and improves the accuracy and richness of the recommendation results.
2) \textbf{Explanability Enhancement.} 
LLM demonstrates strong comprehension and natural language generation capabilities, enabling personalized explanations by effectively understanding user habits and behaviors. Hence, recent research \cite{llmexp1, llmexp2, llmexp3, llmexp4} explores leveraging the LLM for explainable recommendations, providing valuable insights for future work.
3) \textbf{Conversational Enhancement.} LLM's strong language comprehension and generation capabilities are finding extensive application in conversational recommender systems \cite{hecrs, llmconv1, llmconv2}.
Some work treats conversation as a context to enhance LLM's recommendation performance.
Extending LLM-based RS to FL frameworks has intuitive benefits due to the need to collect large amounts of user-side and enterprise-side data.
This architecture fundamentally safeguards the privacy and security of LLM-based RS by exchanging only aggregated model parameters instead of raw sensitive data. 
It can directly leverage the latest edge data from each client to improve the timeliness, accuracy, and personalization of domain-specific user interests for recommendations.

However, the research on LLM-based faces several challenges.
On the one hand, client devices in FedRec are typically personal user devices that lack the resources for fine-tuning and inference, preventing them from benefiting from the LLM.
On the other hand, while enterprises have the computing resources, training the LLM under the FL framework is also challenging to implement due to the protection of data assets.
Based on the parameter type of training or fine-tuning, the existing work of LLM-based can be divided into two types as illustrated in Fig.~\ref{fig:fedllm4rec_emb}: fine-tuning partial parameters of the LLM and fine-tuning external parameters. 

\subsubsection{Fine-tuning Partial Parameters of the LLM}
Fine-tuning some parameters of the LLM within an FL framework is crucial for organizations and companies, as it helps them obtain models that are most adapted to the business side's needs.
However, it is challenging because it requires achieving mutual benefits for multiple parties and ensuring that the companies' data assets are not compromised.
To this end, PPLR \cite{PPLR-zhao2024llm} introduces a feasible elastic training scheme by sharing and collaboratively training non-privacy-sensitive LLM parameters.
This scheme incorporates a dynamic balancing strategy for client-specific parameter aggregation and training and a flexible storage strategy to optimize resource allocation and manage privacy-sensitive parameters.
MRFF \cite{MRRF} improves the multifaceted user modeling capability of foundation model-base RS. The authors capture the sequential patterns of user-item interactions through a self-attention mechanism to realize the multifaceted capture of user interests in FedRec.

\subsubsection{Fine-tuning External Parameters}
Although fine-tuning the LLM shows promising performance, updating the LLM to local devices is impractical in most personalized recommendation scenarios.
In these scenarios, the client is usually a personal user's smartphone, tablet, or computer, which cannot afford fine-tuning and inferencing LLM.
Therefore, a feasible solution for these small clients is to train some external parameters to align with LLM on the client side.
FedPA \cite{FedPA-zhang2024federated} trains lightweight personalized low-rank adapters on client devices using private data. These adapters collaborate with LLM-generated item embeddings received from the server to capture fine-grained user preferences.
FELLAS \cite{yuan2024fellas} sets up an additional LLM server for enriching the textual and sequential representations of item embeddings on the client, transferring knowledge from LLM to the client's sequential recommendation model.

\subsubsection{Open Problems and Future Directions}

Applying LLM to FedRec to improve privacy and recommendation performance is very promising and valuable, yet the overall exploration of this area is still in the early stages.
To this end, we summarize the main challenges for introducing LLM in an FL framework and present potential opportunities for further utility in the FedRec.

\textbf{Explore on cross-device scenarios.} Enabling individual users to benefit from LLM-based FedRec presents significant challenges due to limited client resources (storage, computation, bandwidth) and stricter privacy requirements. Both prompt learning and fine-tuning LLM become considerably more difficult under these constraints, compounded by data sparsity, often resulting in a few-shot learning scenario.
Although FedPA \cite{FedPA-zhang2024federated} and FELLAS \cite{yuan2024fellas} have proposed promising solutions to enable small clients to benefit from LLM, further exploration is still needed. 
For this purpose, we investigate some relevant topics in FL and list them as follows. 
For instance, Wang et al. \cite{llfm-cd1} leverages large-scale global data and the LLM to enable massive parallel federated training and employs distillation techniques to improve the privacy-utility tradeoff. To balance the complexity of LLM and the resource constraints of client devices, Xu et al. \cite{llfm-cd2} uses backpropagation-free training methods, requiring client devices only to execute “perturbed inferences”. Wu et al. \cite{FedBiOT} introduces a resource-efficient approach by generating a lightweight LLM on the server and training a client-side adapter, formulating it as a bi-level optimization problem.

\textbf{More efficient knowledge transfer.} Foundation models like GPT-4 possess extensive knowledge and excel in computer vision and NLP tasks. The training processes on global text data limits their ability to process interactive information and incorporate domain-specific knowledge crucial for recommendation tasks. This hinders the full utilization of the LLM in RS, particularly within the stricter data-sharing constraints of FL frameworks. In FL, Kang et al. \cite{fl-llm-q2-1} provides a comprehensive survey exploring the application of foundation model knowledge to specialized domains. Fan et al. \cite{fl-llm-q2-2} proposes a parameter-efficient federated mutual knowledge transfer framework for mutual enhancement between the server-side LLM and client-side small language models, enabling knowledge exchange and enriching both with domain-specific insights. Additionally, Yu et al. \cite{fl-llm-q2-3} addresses spatial-temporal catastrophic forgetting in the LLM using multi-granularity knowledge representations and prompts for personalized client learning.

\textbf{Decentralized training.} The substantial computational resources required for training and deploying the LLM have become a significant challenge, especially in the FedRec scenario. Recent efforts have expanded the boundaries of the FL framework by proposing decentralized training as a solution. These approaches leverage user-level GPUs and data to enable efficient, lightweight, and privacy-preserving LLM training. Projects like OpenFedLLM \cite{openfedllm} and FederatedScope-LLM \cite{FederatedScope-LLM} focus on designing simple, user-friendly codebases to facilitate decentralized FL training in an accessible and efficient manner. Additionally, Zuo et al. \cite{llmunlearning} introduced a blockchain-based federated LLM framework that leverages blockchain technology to create a tamper-proof record of each model's contributions and innovatively enables efficient unlearning during LLM training within the FL framework.

\textbf{Generative recommendation.} Data sparsity is a common challenge in RS, which is exacerbated in LLM-based FedRec due to strict privacy requirements and limited client data. Leveraging the LLM for data synthesis within FL frameworks has emerged as a promising strategy. For instance, Wang et al. \cite{Self-instruct} proposed a semi-automated self-instruct process that uses instruction signals generated by language models to fine-tune the LLM for specific tasks. Similarly, Zhang et al. \cite{upload} introduces a federated knowledge-transfer loop with a server-side generator producing synthetic data to augment client datasets. Additionally, Wei et al. \cite{simplesynthetic} generates synthetic datasets simulating private client data to create a better initial model, reducing training iterations required for the FL system.

\begin{table*}[t]
\caption{A list of datasets common used in existing FedRec research.}
\centering
\scalebox{0.99}{
\begin{tabular}{|c|c|c|c|} 
\hline
\multicolumn{1}{|c|}{\textbf{Dataset}}           & \multicolumn{1}{c|}{\textbf{Downstream Task}} & \multicolumn{1}{c|}{\textbf{Scenario}} & \multicolumn{1}{c|}{\textbf{Official Link}}  \\
\hline

Amazon review & E-commerce & A/B/C/D  & \url{https://jmcauley.ucsd.edu/data/amazon/}\\ 
\hline
MovieLens & Movie & A/D  & \url{https://grouplens.org/datasets/movielens/} \\
\hline

Yelp & E-commerce & A/B/D  & \url{https://www.yelp.com/dataset} \\
\hline
HetRec & Book, Music, Movie & A  & \url{https://grouplens.org/datasets/hetrec-2011/} \\
\hline

Ciao & Movie & A &  \url{https://www.cse.msu.edu/~tangjili/datasetcode/truststudy.htm}\\
\hline
Epinions & Relation-ship & A &  \url{https://www.cse.msu.edu/~tangjili/datasetcode/truststudy.htm} \\
\hline

Steam & Game & A  & \url{https://cseweb.ucsd.edu/~jmcauley/datasets.html##steam_data} \\
\hline
Yahoo & Movie, Music & A  & \url{https://webscope.sandbox.yahoo.com/catalog.php?datatype=r}
 \\
\hline

MIND & News & A/D  & \url{https://msnews.github.io/}  \\
\hline
Adressa & News & A  & \url{https://reclab.idi.ntnu.no/dataset/}\\
\hline

Taobao & Behavior & A  & \url{https://tianchi.aliyun.com/dataset/649}  \\
\hline
Last.FM & Music & A & \url{https://grouplens.org/datasets/hetrec-2011/}
\\
\hline

Flixster & Relation-ship & A  & \url{https://networkrepository.com/soc-flixster.php} \\
\hline
Gowalla & Check-in & A  & \url{https://www.yongliu.org/datasets/}\\
\hline

Anime& Video & A  & \url{https://www.kaggle.com/datasets/CooperUnion/anime-recommendations-database} \\
\hline
Douban & Movie, Music, Book & A/B/D  & \url{https://www.kaggle.com/datasets/fengzhujoey/douban-datasetratingreviewside-information} \\
\hline

Online-Retail & E-commerce & B  & \url{https://www.kaggle.com/datasets/carrie1/ecommerce-data} \\
\hline
NineRec & General & A/B/C/D  & \url{https://github.com/westlake-repl/NineRec} \\
\hline

Foursquare & Check-in & A/B  & \url{https://paperswithcode.com/dataset/foursquare} \\
\hline
Weeplaces & Check-in & A  & \url{https://www.yongliu.org/datasets/} \\
\hline

ACM & Citation & A  & \url{https://www.aminer.cn/citation} \\
\hline
DBLP & Citation & A  & \url{https://www.aminer.cn/citation} \\
\hline

Tenrec & Video, Article & A/D  & \url{https://tenrec0.github.io/}  \\ 
\hline
KuaiSAR & Video & A/D  & \url{https://kuaisar.github.io/}
 \\
\hline

XING & Job & A  & \url{https://github.com/MilkaLichtblau/xing_dataset}  \\ 
\hline
 Pinterest & Image & A  & \url{https://github.com/edervishaj/pinterest-recsys-dataset}\\

\hline
\multicolumn{4}{|p{\linewidth}|}{\textbf{Scenario:} A: Collaborative FedRec, B: Cross-domain FedRec, C: Multi-modal FedRec, D: LLM-based FedRec.}                 \\
\hline
\end{tabular}
}
\label{tab:Datasets}
\end{table*}

\section{Training and Evaluating}
\subsection{Datasets}

In this section, we introduce the datasets commonly used in FedRec research, summarized in Table~\ref{tab:Datasets}. For convenient reproduction and comparison, we provide the official repository links of the datasets. Following our taxonomy, we summarize the dataset for each scenario as follows:
\begin{itemize}
\item \textit{Collaborative FedRec.} 
This is the most basic recommendation task, so that most datasets can be used for this scenario, researchers and developers can select datasets based on the downstream task.
The MovieLens series is a top choice for benchmarking with multi-scale real movie interaction data (100K/1M/10M/20M, etc.).
However, MovieLens datasets are usually dense, existing research often introduces sparse datasets such as Amazon/Yelp/Last.FM.
They simulate more realistic scenarios and provide technical references for recommendations to the long-tail user groups in the real world.

\item \textit{Cross-domain FedRec.}
Cross-domain modelling heavily relies on multi-platform datasets. The Amazon review dataset, with its multi-scale real heterogeneous transaction data covering categories such as books, electronics, and other goods, serves as the core benchmark in CDR.
It provides a critical scenario for cross-domain feature transfer and distribution differences under the state of data silos in FedRec, effectively supporting the stability verification of cross-domain joint modelling.
    
\item \textit{Multi-modal FedRec.}
Multi-modal recommendation relies on multi-modal feature fusion. NineREC and Amazon Review datasets provide various modal knowledge by covering item images, review texts, and other heterogeneous metadata, serving as benchmarks for multi-modal alignment and knowledge transfer training under privacy protection for FedRec.
    
\item \textit{LLM-based FedRec.}
Although the LLM have the potential for association, inference, retrieval, and generalization, Fedrec's research relies on multi-source heterogeneous data to verify the collaborative efficiency of clients.
Hence, Amazon Review, TenRec, and other enterprise-level datasets, with their coverage of multi-business behavior, become the main benchmarks in this scenario.
    
\end{itemize}

\subsection{Evaluation}

\subsubsection{Strategy}
Most recommendation datasets are explicit datasets due to easier access to user ratings \cite{zhang2014explicit}.
However, they cannot reflect richer users' implicit preferences, such as clicks and browsing.
Hence, some relevant research transforms explicit datasets into implicit ones by transforming ratings higher than 1 to 1 as positive to construct implicit datasets \cite{he2016fast}.
To promote more effective recommendation learning, they randomly sampling several negative items from users' uninteracted items for each positive.
Many works apply the leave-one-out strategy \cite{leave-one-out:deshpande2004item} for evaluation, \textit{e.g.},~holding the last interaction for the test, the second to last for validation, and the others for training. The other works split the original dataset into the training, validation, and test sets with a specific ratio, \textit{e.g.}, 6/3/1, 7/2/1, or 8/1/1.
During the testing process, the testing set is conducted by randomly sampling several (usually 99 or 999) uninteracted items for each positive item.
However, sampling at test time tends to lead to biased recommendation results \cite{li2020sampling}. Therefore, some work also tests all the data not used for training.

\subsubsection{Metric}
FedRec adopts the same evaluation metrics as CenRec.
Widely used metrics evaluate the performance of a FedRec with the top-K predicted positive items contained in the recommendation list. Based on the specific evaluation focus, these metrics can be categorized into three dimensions: \textbf{recommendation accuracy}, \textbf{recommendation list quality}, and \textbf{predicted rating error}. All of the metrics are included in Table~\ref{tab:Metrics}, and we summarize the types of metrics as follows:

\begin{itemize}
    \item \textit{Recommendation accuracy.} This type of metric evaluates the accuracy of the recommendation results. Metrics for this evaluation dimension include: Precision, Recall/ Hit Ratio (HR) and F1 score.
    \item \textit{Recommendation list quality.} This type of metric evaluates whether the recommendation list is optimal. Metrics include: Area Under the Curve (AUC), Mean Average Precision (MAP), Mean Reciprocal Rank (MRR) and Normalized Discounted Cumulative Gain (NDCG).
    \item \textit{Predicted rating error.} This type of metric typically evaluates the correctness of predictive scoring in explicit recommendation tasks. Metrics include: Mean Absolute Error (MAE) and Root Mean Squared Error (RMSE).
\end{itemize}

\begin{table*}
\caption{A list of common metrics used in existing FedRec research.}
\arrayrulecolor{black}
\renewcommand\arraystretch{1.5}
\resizebox{1.0\linewidth}{!}{
\centering
\begin{threeparttable}
\begin{tabular}{|p{1.5cm}|p{1cm}|p{9cm}|p{7.5cm}<{\centering}|} 
\hline

\multicolumn{4}{|p{19cm}|}{ $K$: the length of the recommendation list used for evaluation for a single user. In existing researches, commonly used values for $K$ include $\{1,5, 10, 20, 50\}$.} \\
\multicolumn{4}{|l|}{$\mathcal{K}$: the ranked recommendation list generated in descending order of predicted labels.} \\

\multicolumn{4}{|l|}{$\mathcal{K}_p$: the subset of positive items in $\mathcal{K}$  that the user is genuinely interested in.} \\
\multicolumn{4}{|p{19cm}|}{ $\mathcal{P}_c$: the set of all correctly ranked pairs, where each pair represents a positive item ranked higher than a negative item in $\mathcal{K}$.} \\
\multicolumn{4}{|p{19cm}|}{$\mathcal{K}_{REL}$: the ideal top-$K$ recommendation list ranked by real scores of items in $\mathcal{K}$.} \\
\multicolumn{4}{|p{19cm}|}{$\mathbf{1}(k)$: A binary function indicating whether the item at rank $k$ is positive, \textit{e.g.}, 1 if positive and 0 otherwise.} \\
\multicolumn{4}{|l|}{$\mathcal{I}_t$: the set of positive items for the user in the testing dataset.} \\
\multicolumn{4}{|p{19cm}|}{$rank_p$: represents the position of the first positive item in the ranked list $\mathcal{K}$, $1<rank_p<K$. Specially, when there is no positive item, $rank_p=\infty$.} \\
\multicolumn{4}{|p{19cm}|}{$\mathbf{rel}_k$ represents the real score of the item at rank $k$.} \\
\hline

\multicolumn{1}{|c|}{\textbf{Metric}} & \multicolumn{1}{c|}{\textbf{Type}}          & \multicolumn{1}{c|}{\textbf{Meaning}} & \multicolumn{1}{c|}{\textbf{Formula}} \\
\hline
Precision & A & How many recommended items are truly positive. &

$Precision@K = \frac{|\mathcal{K}_p|}{K}$
\\
\hline
Recall / HR & A & Measure the proportion of truly positive items 
out of $\mathcal{I}_t$. & 
$Recall@K = HR@K = \frac{|\mathcal{K}_p|}{|\mathcal{I}_t|}$ \\
\hline
F1 & A & Evaluate the performance combining Precision and Recall. &
$F1@K=2\times\frac{Precision@K\times Recall@K}{Precision@K + Recall@K}$ \\
\hline
AUC & Q & How well to distinguish between positive and negative
items. & 
$AUC@K = \frac{|\mathcal{P}_c|}{|\mathcal{K}_p|\times(K-|\mathcal{K}_p|)}$ \\
\hline
MAP & Q & How well the system ranks positive items higher in the list. &
$AP@K = \frac{1}{|\mathcal{K}_p|}\sum_{k=1}^K Precision@k\cdot\mathbf{1}(k)$
$MAP@K = \frac{1}{|\mathcal{U}|}\sum^{|\mathcal{U}|}_{u=1}AP^u@K$\\
\hline
MRR & Q & How quickly the
system retrieves a truly positive item for a user. &
$RR@K = \frac{1}{rank_p}$,
$MRR@K = \frac{1}{|\mathcal{U}|}\sum^{|\mathcal{U}|}_{u=1}RR_u@K$
\\
\hline
NDCG & Q & Evaluate
the quality of a ranked list by assigning higher importance to positive items that appear earlier in the list, calculated based on DCG and Ideal DCG (IDCG). &
$DCG@K = \sum_{k=1}^{|\mathcal{K}|}\frac{\mathbf{rel}_k}{\text{log}_2(k+1)}$,
$IDCG@K=\sum_{k=1}^{|\mathcal{K}_{REL}|}\frac{\mathbf{rel}_k}{\text{log}_2(k+1)}$,
$NDCG@K=\frac{DCG@K}{IDCG@K}$\\
\hline
MAE & R & Evaluate the accuracy of predicted ratings. &
$MAE =  \frac{1}{|\mathcal{I}_t|}\sum_{i=1}^{|\mathcal{I}_t|}|\hat{r}_i-r_i|$\\
\hline
RMSE & R & Evaluate the accuracy of predicted ratings. &
$MSE =  \frac{1}{|\mathcal{I}_t|}\sum_{i=1}^{|\mathcal{I}_t|}(\hat{r}_i-r_i)^2$,
$RMSE = \sqrt{MSE} $\\
\hline
\multicolumn{4}{|p{19cm}|}{\textbf{Type:} A: Recommendation accuracy, Q: Recommendation list quality, R: Predicted rating error.} \\
\hline
\end{tabular}
\end{threeparttable}
}
\label{tab:Metrics}
\arrayrulecolor{black}
\end{table*}

\section{Challenge and Future Direction}

In the previous chapters, we have discussed the unique challenges and open issues of each scenario. Below we discuss the common challenges and potential opportunities in the whole FedRec field.

\subsection{Trustworthy}
With the continuous development of human-centred artificial intelligence, more and more work on trustworthy artificial intelligence especially for RS has emerged \cite{ge2024survey}. Current research in FedRec predominantly focuses on validating the effectiveness of distributed architectures, while mainly overlooking pivotal demands in \textit{fairness} and \textit{explainability}. 

\textit{Fairness.} In addition to the inherent biases of the recommendation \cite{PAMI_bias}, FedRec confronts dual fairness challenges. On the one hand, the compounding effects of bias exhibited in recommender systems amplify group discrimination risks \cite{wang2023survey}.
Specifically, interaction data exhibits statistical shifts as long-tail user sparsity and Matthew effects in item exposure \cite{jiang2024item}. Additionally, algorithms suffer from gradient biases as collaboration mechanisms tend to favor active users \cite{wang2024counterfactual, greenwood2024user}. Both data-level and algorithm-level biases create a positive feedback loop that systematically neglects disadvantaged user groups, \textit{e.g.}, ethnic minorities, low-engagement users, etc.. While F$^2$PGNN \cite{F2PGNN-agrawal2024no} pioneers population bias mitigation via graph neural network-based group-aware modeling, academic and industrial scenarios need more universal fairness frameworks.
On the other hand, there is an inherent dilemma of client participation fairness in the FL architecture.
Current federated training protocols, prioritizing global objective through client selecting and gradient weighting strategies, naturally filtering the exclusion of participants with inferior device capabilities or unique data distributions \cite{mohri2019agnostic, lifair}.

\textit{Explainability.} The explainability of RS significantly influence the user trust and acceptance of suggestions \cite{TPAMI_disentangled}. In industrial applications, demands of explainability exhibit significant branches. On user-level, the explainable RS generates rationale on why system recommends this item, affecting the user immediate decision \cite{expl-zhang2020explainable,expl-geng2022path,expl-li2024attention, expl-ma2023kr}. Yet on enterprise-side, it can help verify and optimize the logic of traffic allocation. Predominant black-box optimization paradigms in FedRec fail to reveal recommendation reasons, thus becoming increasingly inadequate under regulatory pressures like DSA\footnote{\url{https://commission.europa.eu/strategy-and-policy/priorities-2019-2024/europe-fit-digital-age/digital-services-act_en}}. Promising breakthroughs in the future include: (1) Federated explainability middleware, which enables encrypted-domain recommendation traceability via cross-platform item association analysis; (2) Privacy-preserving explanation protocols, which generates natural language interpretations from distributed user profiles without exposing raw features.

\subsection{Efficient Training}

Research in collaborative FedRec has already pay much attention on efficient training. The reason comes from that inferior user devices are unable to afford original model training and sharing. As discussed in Section \ref{Sec:facilitate}, communication optimizations often involve model compression techniques \cite{LightFR-zhang2023lightfr, PO-FCF-khan2021payload, CoLR-nguyen2024towards} or parameter generation methods \cite{MetaMF-lin2020meta, ReFRS-imran2023refrs}. Computation optimizations include adaptive model size adjustments \cite{RF2-maeng2022towards, Efficient-FedRe-yi2021efficient, HeteFedRec-yuan2024hetefedrec} to reduce the inference burden on clients.

However, there has been little research on context-enhanced scenarios widely utilized in industry. Compared with the collaboration-only system, although clients usually hold stronger capability in computing and communicating, they need to process more large-scale and heterogeneous data, and typically integrate more complicated architectures. For example, in multi-modal FedRec, content understanding across short video platforms needs to pay much more effort on exact image or text modalities than collaborative information. Methods like AMMFRS \cite{AMMFRS-feng2024recommendation} which settles encoders like ResNet or BERT on inferior user devices in theory for promising performance, which actually cannot be achieved in deployment. Similar dilemmas also exist in LLM-based FedRec like cross-lingual recommendations in cross-border e-commerce.

In context-enhanced FedRec deployment, client capability and scenario resource requirements scale proportionally. Vanilla methods suffer from collaboration redundancy, highlighting the significance of efficient training. Yet it remains an overlooked area.
This gap highlights a critical turning point in the evolution of FedRec, especially in context-enhanced scenarios. Efficient management of computation, storage, and communication resources on client devices must be deeply integrated with the requirements of specific business applications.

\subsection{User-governed Learning}

Existing research on FedRec predominantly operates under the assumption of a uniformly scheduling framework, which thoroughly overlooks the pivotal role of users as autonomous decision-makers in recommendation processes. Two critical dimensions on privacy warrant reflection. First, the imperative of "the right to be forgotten" (RTBF) has gained increasing attention \cite{UL-xu2024machine}. Individuals who contribute their preferences to enable recommendation services ought to retain the right to withdraw from FL frameworks, and demand the obliteration of all preference-related affects from system memory. Concurrently, system organizers should be responsible for robust architectures capable to mitigate malicious client intrusions and recovering from data poisoning attacks \cite{revisit-su2024revisit, rong2022fedrecattack}. While FRU \cite{FRU-yuan2023federated} pioneers the integration of machine unlearning into FedRec to eliminate client data retention, unresolved challenges persist. Concerns include addressing model heterogeneity during dynamic user withdrawal and establishing reliable verification mechanisms for confirming the certified removal of either the entire client or partial client-specific data \cite{UL-liu2024survey}. Second, user must extend to fine-grained control over the personal impression left to the RS.
In certain cases, users may seek to construct a virtual persona to conceal their true preferences from the recommendation system.
CDCGNNFed \cite{CDCGNNFed-qu2024towards} proposes a user-governed interactions sharing mechanism, yet the system can still infer users’ true preferences from their sampled interaction data. Such issues are prevalent in cross-domain scenarios where users may seek to construct different digital personas across domain, which contradicts the existing alignment protocols. In addition to privacy concern, user-centric recommender systems have inspired new thinking about user-governed learning. In addition to ensuring traditional recommendation accuracy, how can FedRec meet the need for diversity \cite{UL-coppolillo2024relevance, UL-zheng2024diversity} in decentralized environments? These imperatives necessitate novel architectures for user-governed learning.

\subsection{Unified Benchmark}
In CenRec, prominent frameworks such as RecBole \cite{RecBole-zhao2021recbole, RecBole2-zhao2022recbole} and FuxiCTR \cite{CTR-zhu2021open, CTR-zhu2022bars} have systematically developed well-established benchmarks for fundamental experimental configurations and performance comparisons \cite{TPAMI_benchmark}.  However, current public code repositories for FedRec research exhibit significant discrepancies in their implementations. They utilize divergent machine learning frameworks, algorithm libraries, training configurations, and dataset partitioning strategies. These disparities pose substantial challenges to baseline reproduction and objective model evaluation. Apart from that, classic recommendations are commonly transferred to FL as baselines, \textit{e.g.}, MF \cite{FedHGNN-yan2024federated, FedRAP-lifederated}, NCF \cite{FedNCF-perifanis2022federated,PFedRec-zhang2023dual}, DCN \cite{IFedRec-zhang2024federated}, and LightGCN \cite{GPFedRec-zhang2024gpfedrec, PTF-FedRec-yuan2024hide}. Existing open-source projects yet lack unified deployment protocols and standardized training procedures of them. As a result, there is a pressing need to establish specialized FedRec libraries and a comprehensive benchmark. This benchmark should enable fair comparisons across multiple dimensions including accuracy, resource consumption, fairness, exlainability, and other critical criteria. FL benchmarks as PFLlib \cite{PFLlib-zhang2023pfllib} can provide instructions for this direction. Additionally, a feasible FedRec benchmark should comprehensively encompass diverse task scenarios, spanning single-domain and cross-domain applications, cross-device and cross-silo architectures, as well as unimodal and multimodal learning paradigms. The interoperability across multiple scenarios will bring more application directions for the deployment of FedRec.

\subsection{Online Federated Recommendation}

Current research in FedRec predominantly operates under static scenario assumptions, where items and users are presumed known in advance. However, practical recommendation applications face critical challenges on dynamic changes, manifesting in three dimensions.
First, recommendation often need to integrate with other services like advertisements in real time. This necessitates that FedRec support dynamic cross-service knowledge transfer \cite{IncMSR-zhang2024incmsr}. Second, user interests and item popularity are constantly evolving. Some part of individual preference fades over time while new parts emerge \cite{IMSR-wang2023incremental, Graphsail-xu2020graphsail, INFER-zhang2024influential}. Moreover, for time-sensitive items like news or digital products, their characteristics can change significantly over time. Third, new users and items are continuously entering the system \cite{FIRE-xia2022fire, IMA-wang2024framework}. These new entities lack collaborative information, requiring effective online cold-start strategies. As data volumes grow exponentially, static models relying on fixed data distributions inevitably suffer from performance degradation \cite{FIL-li2024towards, GLFC-dong2022federated}. While full-parameter retraining mitigates this issue, frequent iterations of the offline training to online deployment cycle result in service downtime and commercial revenue loss \cite{LRURec-yue2024linear, LSAT-shi2024preliminary}, whether for lightweight models or the LLM. This remains a core bottleneck constraining broadly applications for FedRec. Consequently, establishing a continuous federated evolution mechanism grounded in incremental learning has become imperative for practical deployment \cite{FedMKL-m2022personalized}. There remains an urgent need to develop online frameworks for FedRec that harmonize privacy-preserving real-time learning with dynamic trade-offs among communication efficiency, computational complexity, and recommendation quality. 

\subsection{Multiple Attack and Defense}
While FedRec mitigate privacy risks without directly sharing data, indirect privacy leakage remains a critical focus. Senior resolutions like differential privacy or homomorphic encryption exhibit performance degradation against organized attacks, leading research to predominantly focus on malicious client-induced threats such as model poisoning \cite{ConDA-liang2025defending, PoisonFRS-yin2024poisoning}, untargeted attacks \cite{HiAttack-ali2024hidattack}, and membership inference attacks \cite{CIRDP-liu2024defending}. However, current studies suffer from limitations. First, excessive emphasis on federated architecture-derived attack scenarios overlooks the intrinsic requirements on defense to privacy inference in recommender systems \cite{ali2025privacy}. Compared with the degradation on recommendation quality, unauthorized inference of user privacy fundamentally contradicts the basic intent of federated mechanisms. Second, in cross-platforms scenarios like the cross-domain FedRec, private attributes of the user remain susceptible to inverse inference via multi-platform feature correlation. This reveals the shortcoming of current architectures on cross-domain isolation mechanisms. In a word, the fundamental contradiction between current attack-defense research and application requirements lies in the fact that, in real-world scenarios, attack methods often exhibit polymorphous characteristics. In industrial context-enhanced applications, attack paths dynamically evolve with features of the enhancement route. As a result, it remains valuable to establish dynamic defense frameworks which get round with the multi-attack threat and exhibit adaptability to different scenarios for advancing FedRec.

\section{Conclusion}
Federated recommender system (FedRec) has attracted academic attention because they efficiently protect the privacy of users and organizations without compromising performance.
Previous reviews analyze and explain how to transfer general recommender systems (RS) to federated learning (FL) architectures from the perspective of FL scholars, ignoring the unique properties and practical challenges of fine-grained scenarios.
To fill this gap, this paper provides a systematic and in-depth review of extending modern RS to the FL framework from the perspective of recommendation researchers and practitioners.
Compared to previous reviews, we comprehensively analyze the coupling of recommender systems and federated learning, providing a more practical insight into the application.
We aim to develop technology guidance for FedRec, filling the gap between existing reviews and real deployment.
Specifically, we establish clear connections between fine-grained scenarios and FL frameworks, including collaborative FedRec, cross-domain FedRec, multi-modal FedRec, and LLM-based FedRec.
This review comprehensively discusses the work and key challenges in these critical scenarios and suggests potential opportunities. For those early-stage scenarios, we offer potential solutions and directions for development in conjunction with related FL topics.
Furthermore, we explore and point out common issues and promising directions for the entire FedRec field to provide a forward-looking perspective.
Overall, this paper pioneers in encouraging the deployment of FedRec and interdisciplinary research. It promotes the comprehensive deployment of privacy-preserving recommendations under data compliance constraints.

\section{Limitations}
Although we have tried our best to clarify the development status, challenges and opportunities of FedRec, there are still some limitations.
Firstly, due to page limitations, we can only show an overview of some key FedRec scenarios rather than every segmented task and technical detail.
Secondly, due to Fedrec's rapid development, this review mainly covers key research from conferences as WWW, KDD, SIGIR, ICLR, IJCAI, NeurIPS, AAAI, journals as TOIS and TKDE, and arXiv up to early 2025. We will stay updated with the community's development and plan to continuously revise our work in the future.

\bibliographystyle{IEEEtran}
\bibliography{sample-base}

\end{document}